\documentclass{msc-modified}
\usepackage{lineno}
\nolinenumbers
\usepackage[british]{babel}

\usepackage[leqno]{amsmath}
\usepackage{stmaryrd}
\usepackage{amsfonts}
\usepackage{mathtools}
\usepackage{amsthm}%
\usepackage{thmtools,thm-restate}
\usepackage{bm}%
\usepackage{array}

\usepackage{paralist}
\usepackage{multirow}
\usepackage{booktabs}%

\usepackage{tikz}
\usetikzlibrary{decorations.pathreplacing}

\usepackage{xspace}

\usepackage{xcolor}
\definecolor{Prune}{RGB}{99,0,60}
\definecolor{B1}{RGB}{49,62,72}
\definecolor{C1}{RGB}{124,135,143}
\definecolor{D1}{RGB}{213,218,223}
\definecolor{A2}{RGB}{198,11,70}
\definecolor{B2}{RGB}{237,20,91}
\definecolor{C2}{RGB}{238,52,35}
\definecolor{D2}{RGB}{243,115,32}
\definecolor{A3}{RGB}{124,42,144}
\definecolor{B3}{RGB}{125,106,175}
\definecolor{C3}{RGB}{198,103,29}
\definecolor{D3}{RGB}{254,188,24}
\definecolor{A4}{RGB}{0,78,125}
\definecolor{B4}{RGB}{14,135,201}
\definecolor{C4}{RGB}{0,148,181}
\definecolor{D4}{RGB}{70,195,210}
\definecolor{A5}{RGB}{0,128,122}
\definecolor{B5}{RGB}{64,183,105}
\definecolor{C5}{RGB}{140,198,62}
\definecolor{D5}{RGB}{213,223,61}

\usepackage{hyperref}
\usepackage{natbib}%
\usepackage[capitalise,noabbrev,nameinlink]{cleveref}
\usepackage[cleveref,hyperref,electronic,xcolor]{knowledge}
\knowledgeconfigure{notion}
\hypersetup{
  hypertexnames=false,
	colorlinks=true, %
	urlcolor=A5, %
	linkcolor=A4, %
	citecolor=Prune
}

\knowledgestyle{intro notion}{color={A4}, emphasize}
\knowledgestyle{notion}{color={B1}}

\newcommand{\defined}{\stackrel{\mbox{\begin{scriptsize}\textsf{def}\end{scriptsize}}}{=}}
\newcommand{\eqdef}{\stackrel{\mbox{\begin{scriptsize}\textsf{def}\end{scriptsize}}}{=}}
\newcommand{\equivdef}{\stackrel{\mbox{\begin{scriptsize}\textsf{def}\end{scriptsize}}}{\iff}}
\newcommand{\eqby}[1]{\stackrel{\mbox{\begin{scriptsize}\!#1\!\end{scriptsize}}}{=}}

\newcommand{\unknown}{\textcolor{C3}{\textbf{?}}}

\newcommand{\set}[1]{\{\; {#1} \;\}}
\newcommand{\mset}[1]{\{\!|\; {#1} \;|\!\}}
\newcommand{\setof}[2]{\{\; {#1} \mid {#2} \;\}}
\newcommand{\Setof}[2]{\bigl\{\; {#1} \big| {#2} \;\bigr\}}

\newcommand{\Nat}{\mathbb{N}}

\newcommand{\Bad}{\mathop{\mathsf{Bad}}}
\renewcommand{\Dec}{\mathop{\mathsf{Dec}}}%
\newcommand{\Inc}{\mathop{\mathsf{Inco}}}%
\newcommand\Down{\mathop{\mathsf{Down}}\nolimits}
\newcommand{\Idl}{\mathop{\kl[idl-completion]{\mathsf{Idl}}}}
\knowledge{idl-completion}{notion}
\newcommand\Ant{\operatorname{\kl[antichains-set]{\mathsf{Ant}}}}
\knowledge{antichains-set}{notion}

\newcommand{\FAnt}[1]{\operatorname{\kl[antichains-set-finite]{\Gamma}}_{#1}}
\knowledge{antichains-set-finite}{notion}

\newcommand{\BadSeq}{\mathop{\kl[BadSeq]{\mathsf{Bad}}}}
\newcommand{\DecSeq}{\mathop{\kl[DecSeq]{\mathsf{Dec}}}}
\newcommand{\IncSeq}{\mathop{\kl[IncSeq]{\mathsf{Inco}}}}
\knowledge{BadSeq}{notion}
\knowledge{DecSeq}{notion}
\knowledge{IncSeq}{notion}

\newcommand{\dunion}{\mathop{\kl[dunion]{\sqcup}}}
\newcommand{\Dunion}{\mathop{\kl[dunion]{\bigsqcup}}}
\knowledge{dunion}{notion}

\newcommand{\lexsum}{\mathop{\kl[lexsum]{+}}}
\newcommand{\Lexsum}{\mathop{\kl[lexsum]{\sum}}}
\knowledge{lexsum}{notion}

\newcommand{\cartprod}{\mathop{\kl[cartprod]{\times}}}

\knowledge{cartprod}{notion}

\newcommand{\lexprod}{\mathop{\kl[lexprod]{\cdot}}}

\knowledge{lexprod}{notion}

\newcommand\dc{\mathop{\downarrow}\nolimits}
\newcommand\dwc{\mathop{\kl[dwc]{\downarrow}}\nolimits}
\knowledge{dwc}{notion}

\newcommand{\DefineNewOrder}[4]{%
  \expandafter\newcommand\csname#1le\endcsname%
    {\mathrel{\kl[#1-ordering]{{#2}_{#4}}}}%
	\expandafter\newcommand\csname#1ge\endcsname%
    {\mathrel{\kl[#1-ordering]{\reflectbox{\text{$#2$}}}_{#4}}}%
	\expandafter\newcommand\csname#1leq\endcsname
    {\mathrel{\kl[#1-ordering]{{#3}_{#4}}}}%
	\expandafter\newcommand\csname#1geq\endcsname
    {\mathrel{\kl[#1-ordering]{{\text{\reflectbox{$#3$}}}_{#4}}}}%
	\expandafter\newcommand\csname#1eq\endcsname
    {\mathrel{\kl[#1-ordering]{{\equiv}_{#4}}}}%
	\expandafter\newcommand\csname#1orth\endcsname
    {\mathrel{\kl[#1-ordering]{{\bot}_{#4}}}}%
  \knowledge{#1-ordering}{notion}%
}

\newcommand{\leqaug}{\mathrel{\kl[leqaug]{\leq_{\mathsf{aug}}}}}
\knowledge{leqaug}{notion}
\newcommand{\leqstruct}{\mathrel{\kl[leqstruct]{\leq_{\mathsf{st}}}}}
\knowledge{leqstruct}{notion}

\newcommand{\leqapprox}{\mathrel{\kl[leqapprox]{{<}_{\mathsf{approx}}~}}}
\knowledge{leqapprox}{notion}

\newcommand{\refl}{\mathop{\kl[reflectsin]{\hookrightarrow}}}%
\knowledge{reflectsin}{notion}

\newcommand{\geqaug}{\geq_{\mathsf{aug}}}
\newcommand{\geqstruct}{\geq_{\mathsf{st}}}

\newcommand{\leqcm}{\mathrel{\kl[leqcm]{\leq_{\mathsf{cond}}}}}
\knowledge{leqcm}{notion}

\newcommand{\leqcmstruct}{\mathrel{\kl[leqcmstruct]{\leq_{\mathsf{cond+st}}}}}
\knowledge{leqcmstruct}{notion}

\newcommand{\wpoiso}{\mathrel{\kl[wpoiso]{\cong_{\mathsf{wpo}}}}}
\knowledge{wpoiso}{notion}

\newcommand{\orank}{\operatorname{\kl[orank]{rk}}}
\knowledge{orank}{notion}

\newcommand{\wqo}{\ensuremath{\kl[well-quasi-order]{\mathsf{wqo}}}\xspace}
\newcommand{\wpo}{\ensuremath{\mathsf{wpo}}\xspace}
\newcommand{\wqos}{\ensuremath{\mathsf{wqos}}\xspace}
\newcommand{\wpos}{\ensuremath{\mathsf{wpos}}\xspace}

\newcommand\om{\omega}
\newcommand{\omom}{{\om^{\om}}}

\newcommand\Hoare{\mathcal H}
\DefineNewOrder{hoare}{\sqsubset}{\sqsubseteq}{\Hoare}

\newcommand\Smyth{\mathcal S}
\DefineNewOrder{smyth}{\sqsubset}{\sqsubseteq}{\Smyth}

\newcommand\Pf{\operatorname{\kl[finite-powerset]{\mathsf{P_f}}}}
\knowledge{finite-powerset}{notion}
\newcommand\Pfull{\operatorname{\kl[infinite-powerset]{\mathsf{P}}}}
\knowledge{infinite-powerset}{notion}

\newcommand{\Mul}{\mathop{\kl[multisets]{\mathsf{M^{\diamond}}}}}
\knowledge{multisets}{notion}
\newcommand{\Muln}{\mathop{\kl[multiset-n]{\mathsf{M_n^{\diamond}}}}}
\newcommand{\Multwo}{\mathop{\kl[multiset-n]{\mathsf{M_2^{\diamond}}}}}
\knowledge{multiset-n}{notion}

\DefineNewOrder{mult}{<}{\leq}{\diamond}
\newcommand{\seq}{{<\omega}}

\newcommand{\someInvariant}{\mathop{\mu}}
\newcommand{\someRelation}{\mathrel{\bullet}}

\newcommand{\h}{\operatorname{\mathbf{\kl[ordinal-height]{h}}}}
\newcommand{\w}{\operatorname{\mathbf{\kl[ordinal-width]{w}}}}
\renewcommand{\o}{\operatorname{\mathbf{\kl[ordinal-type]{o}}}}

\newcommand{\fot}{\o_{\bot}}
\newcommand{\oinf}{\operatorname{\kl[approx-ord]{\underline{\mathbf{o}}}}}
\newcommand{\hinf}{\operatorname{\kl[approx-ord]{\underline{\mathbf{h}}}}}
\newcommand{\winf}{\operatorname{\kl[approx-ord]{\underline{\mathbf{w}}}}}
\newcommand{\someApproxInvariant}{\operatorname{\kl[approx-ord]{\underline{\mu}}}}
\knowledge{approx-ord}{notion}

\newcommand{\IncPairs}[1]{\operatorname{\kl[increasing-pairs]{\mathsf{Inc}_{#1}}}}
\knowledge{increasing-pairs}{notion}

\newcommand{\nadd}{\oplus}

\newcommand{\osup}[2]{\sup_{#2} \left\{{#1}\right\}}

\newcommand{\hplus}{\mathbin{\hat{\oplus}}}

\newcommand{\oprim}[1]{#1^{\circ}}
\newcommand{\omprim}[1]{{\om}^{\oprim{#1}}}

\newcommand{\hprod}{\odot}

\newcommand{\FSim}[1]{\mathcal{P}_{#1}}

\newcommand{\FPhi}[1]{\mathcal{H}_{#1}}

\crefname{therm}{Theorem}{Theorems}
\Crefname{therm}{Theorem}{Theorems}

\makeatletter%
\def\cleartheorem#1{%
    \expandafter\let\csname#1\endcsname\relax
    \expandafter\let\csname c@#1\endcsname\relax
}
\makeatother

\theoremstyle{plain}
\cleartheorem{therm}
\newtheorem{therm}{Theorem}[section]
\cleartheorem{lemma}
\newtheorem{lemma}[therm]{Lemma}
\cleartheorem{fact}
\newtheorem{fact}[therm]{Fact}
\cleartheorem{proposition}
\newtheorem{proposition}[therm]{Proposition}
\cleartheorem{corollary}
\newtheorem{corollary}[therm]{Corollary}
\cleartheorem{example}
\newtheorem{example}[therm]{Example}
\theoremstyle{definition}
\cleartheorem{definition}
\newtheorem{definition}[therm]{Definition}
\cleartheorem{remark}
\newtheorem{remark}[therm]{Remark}

\knowledge{notion}
  | well-quasi-order

\knowledge{notion}
  | maximal order type
  | maximal order types
  | m.o.t.

\knowledge{notion}
  | ordinal height
  | height

\knowledge{notion}
  | ordinal width
  | width
  | widths@ordinal
  | width@ordinal

\knowledge{notion}
  | ordinal invariants
 | ordinal invariant

\knowledge{notion}
  | ordinal-type
\knowledge{notion}
  | ordinal-height
\knowledge{notion}
  | ordinal-width

\knowledge{notion}
 | elementary \wqos 
 | elementary \wqo 
 | elementary $\wqo $
 | elementary

\knowledge{notion}
 | $\omega $-elementary \wqos 
 | $\om $-elementary \wqo 

\knowledge{notion}
  | good
  | good sequences

\knowledge{notion}
  | bad sequences
  | bad

\knowledge{notion}
  | finite antichain condition

\knowledge{notion}
 | antichain
 | antichains

\knowledge{notion}
  | well-partial-order

\knowledge{notion}
  | {\wpo }-isomorphic
  | \wpo -isomorphism
  | \wpo -isomorphic
  | {\wpo }-isomorphisms

\knowledge{notion}
  | lexicographic sum
  | lexicographic sums

\knowledge{notion}
  | Cartesian product
  | Cartesian products

\knowledge{notion}
  | lexicographic product
  | lexicographically ordered pairs

\knowledge{notion}
  | words
 | finitary sequence
  | finite sequence
\knowledge{notion}
  | word embedding relation
  | embedding@word

\knowledge{notion}
  | set of finite multisets
 | finitary multiset
 | finite multisets
 | finite multiset

\knowledge{notion}
  | multiset embedding

\knowledge{notion}
  | Hoare embedding
  | domination quasi-ordering
  | Hoare's embedding
  | lower preorder

\knowledge{notion}
  | $\varepsilon $-number

\knowledge{notion}
  | additively indecomposable

\knowledge{notion}
  | multiplicatively indecomposable
  | (multiplicatively) indecomposable
  | indecomposable

\knowledge{notion}
  | ordinal rank
 | rank function
  | rank of $T$
  | rank

\knowledge{notion}
  | well-founded quasi-orders
  | well-founded

\knowledge{notion}
  | descent-equations
  | descent equations
  | descent equation

\knowledge{notion}
  | residuals
  | residual

\knowledge{notion}
  | downward closure
 | downward-closure
\knowledge{notion}
 | downwards closed
 | downwards-closed
\knowledge{notion}
 | up-directed
\knowledge{notion}
 | ideals
 | ideal
\knowledge{notion}
 | ideal completion

\knowledge{notion}
  | finitary powerset
 | finitary powersets
  | finitary powerset construction

\knowledge{notion}
  | reflections
  | reflects into
  | reflection

\knowledge{notion}
  | substructures
  | substructure

\knowledge{notion}
  | augmentations
  | augmentation
\knowledge{notion}
 | condensations
 | condensation

\knowledge{notion}
  | disjoint unions
  | disjoint sum
  | disjoint sums

\knowledge{notion}
 | CNF
 | Cantor normal form
 | Cantor Normal Form

\knowledge{notion}
 | approximated \wqos 
 | approximates
 | approximation
 | approximations

\knowledge{notion}
 | weakened m.o.t.
 | weakened width
 | weakened maximal order type
 | weakened ordinal invariant

\begin{document}

\title{Measuring well-quasi-ordered finitary powersets}
\lefttitle{Measuring well-quasi-ordered finitary powersets}
\righttitle{Mathematical Structures in Computer Science}

\papertitle{Article}

\jnlPage{1}{00}
\jnlDoiYr{2023}
\doival{10.1017/xxxxx}

\begin{authgrp}
\author{Sergio Abriola\textsuperscript{1,2}}
\author{Simon Halfon\textsuperscript{3}}
\author{Aliaume Lopez\textsuperscript{4}}
\author{Sylvain Schmitz\textsuperscript{5}}
\author{Philippe Schnoebelen\textsuperscript{3}}
\author{~Isa Vialard\textsuperscript{3}}
\affiliation{\textsuperscript{1}
Universidad de Buenos Aires, Facultad de Ciencias Exactas y Naturales, Departamento
de Computación, Buenos Aires, Argentina}
\affiliation{\textsuperscript{2}
CONICET-Universidad de Buenos Aires, Instituto de Ciencias de la Computación (ICC), Buenos Aires, Argentina}
\affiliation{\textsuperscript{3}
Université Paris-Saclay, CNRS, ENS Paris-Saclay, Laboratoire Méthodes Formelles (LMF), Gif-sur-Yvette, France}
\affiliation{\textsuperscript{4}
Univ.\ Bordeaux, CNRS, Bordeaux INP, LaBRI, Talence, France}
\affiliation{\textsuperscript{5}
Université Paris Cité, CNRS, IRIF, Paris, France}

\end{authgrp}

\begin{abstract}
The complexity of a well-quasi-order (\wqo) can be measured through
three ordinal invariants: the width as a measure of
antichains, height as a measure of chains, and maximal order type as
a measure of bad sequences.

We study these ordinal
invariants for  the finitary powerset, i.e., the collection
$\Pf(A)$ of finite subsets of a \wqo $A$ ordered with the Hoare
embedding relation.   We show that the invariants of $\Pf(A)$ cannot be expressed as a
function of the invariants of $A$, and provide tight upper and lower
bounds for them.

We then focus on a family of well-behaved \wqos, for which these
invariants can be computed compositionally, using a newly defined
ordinal invariant called the \emph{approximate} maximal order type.
This family  is built from multiplicatively
indecomposable ordinals, using classical operations such as disjoint
unions, products, finite words, finite multisets,
and  the finitary powerset construction.
\end{abstract}

\maketitle

\section{Introduction}
\label{sec:introduction}

\AP
Well-quasi-orders (\wqos) are an extensively used and studied tool in
mathematics, logic and computer science \citep[see, e.g., the recent book by][]{schuster2020}, with a wide range of
applications,  notably
in graph
theory~\citep{robertson2004} and in our own specific field of interest, the automatic verification of computer
programs and
systems~\citep{dershowitz79,abdulla2000,finkel2001,blass2008,anand2024}.%

\AP
For a finite partial order (which is always a \wqo), several natural invariants
exist: its size, its ``height'', and its ``width'', as illustrated in
\cref{fig:hasse-width-height-size}.
Analogues of these measures also exist for
general (i.e. infinite) well-quasi-orders. The \reintro{maximal order type} (or
m.o.t.) was originally defined by \cite{dejongh77} as the order type of the
maximal linearisation of a \wqo. \cite{schmidt81} then introduced the
\reintro{ordinal height} as the order type of a maximal chain of a \wqo. Later
\cite{kriz90b} crafted the notion of \reintro{ordinal width}, standardized the
definitions of these three \reintro{ordinal invariants}, and proved numerical
relationships between them (see, for instance,  \Cref{thm:kt} and its
corollary). 
These invariants can be used to compare the ``combinatorial complexity'' of \wqos, following a terminology from \citet{selivanov2025}.

\begin{figure}
  \centering
  \begin{tikzpicture}
    \node[inner sep=1pt,fill=black,circle,draw] (a) at (0,0) {};
    \node[inner sep=1pt,fill=black,circle,draw] (b) at (-1,1) {};
    \node[inner sep=1pt,fill=black,circle,draw] (c) at (0,1) {};
    \node[inner sep=1pt,fill=black,circle,draw] (d) at (1,1) {};
    \node[inner sep=1pt,fill=black,circle,draw] (e) at (0,2) {};
    \node[inner sep=1pt,fill=black,circle,draw] (f) at (0,3) {};
    \draw (a) -- (b);
    \draw (a) -- (c);
    \draw (a) -- (d);
    \draw (b) -- (e);
    \draw (c) -- (e);
    \draw (d) -- (e);
    \draw (e) -- (f);

    \draw[decorate,decoration={brace,amplitude=10pt}] (1.3,0) -- (-1.3,0)
      node[midway,yshift=-15pt] {\footnotesize width = 3};
    \draw[decorate,decoration={brace,amplitude=10pt,mirror}] (-1.5,2.8) -- (-1.5,0.2)
      node[midway,xshift=-15pt, rotate=90] {\footnotesize height = 4};
  \end{tikzpicture}
  \caption{A partial order of size 6, height 4 and width 3, described
  by its Hasse diagram.}
  \label{fig:hasse-width-height-size}
\end{figure}

For well-structured transition systems, i.e., computational systems
that rely on an underlying \wqo
\citep{abdulla2000,finkel2001,bonnet2013}, one can assign
complexity upper bounds related to maximal order types of \wqos through the technology
of \emph{length function theorems} and \emph{controlled bad sequences}
\citep[e.g.,][]{FFSS-lics2011,abriola2015,balasubramanian2020}.
\cite{schmitz2019b} refined this technique by using \emph{controlled
antichains} to prove complexity upper bounds related to widths of \wqos instead of maximal order types.
In all these examples, the computational complexity bounds are derived from bounds on the
ordinal invariants of the underlying \wqos.

When, as is often the case in programming, a \wqo is obtained by combining simpler \wqos,
computing its ordinal invariants can often be done compositionally since
usually these invariants are expressible as a function of the invariants of the
components.
\Citeauthor{dejongh77}
initiated this line of thinking by computing the m.o.t.\ of  disjoint sums and
Cartesian products of \wqos \citep{dejongh77}, followed by
\citeauthor{schmidt2020} with the m.o.t.\ of word embedding and
homeomorphic tree embedding on a \wqo \citep{schmidt2020} and
\citeauthor*{vandermeeren2015} with the
m.o.t.\ of the finite multiset construction
\citep{vandermeeren2015,weiermann2009}.
Considering the other invariants, \citeauthor{abbo99} measured the height of
Cartesian products, and also the width of disjoint sums and
lexicographic products \citep{abbo99}.  We refer to \citep{dzamonja2020}
for a recent survey of these questions.

However, there are useful operations on \wqos for which computing the ordinal invariants
remains a challenge. For instance, the width of the
Cartesian product $A\times B$ of two \wqos is not a function of the ordinal
invariants of $A$ and $B$, as shown by \citeauthor{vialard2024} who
tackled that issue by exhibiting a
family of \emph{elementary \wqos}, for which the width of Cartesian products
is compositionally computable \citep{vialard2024}. Similarly, the
width of the multiset ordering with elements from $A$ cannot be
expressed as a function of $A$'s width, height and m.o.t., and
\citet{vialard2023} proved it is equal to $\omega^{\fot(A)}$,
introducing a new ordinal invariant, $\fot(A)$, the \emph{friendly order type} of $A$.

This article focuses on another difficult case, namely the finitary powerset
construction, i.e., the set  $\Pf(A)$  of all \emph{finite} subsets of some
\wqo $A$, ordered by embedding (unlike the full powerset, this is a \wqo
whenever $A$ is a \wqo).

\paragraph*{Contributions.} Our contributions are threefold. First, we
establish lower and upper bounds on the invariants of $\Pf(A)$ with respect to
those of $A$, as summarized in \Cref{table:ordinal-invariants-pf}. Second, we
prove that these bounds are tight by constructing two families of \wqos,
$\FSim{}$ and $\FPhi{}$ reaching those bounds. Finally, noticing that the
examples witnessing the tightness of the bounds satisfy the peculiar property
that $\o(A)=\w(A)$, and because our bounds prove that $\o(\Pf(A))=\w(\Pf(A))$
in such cases, we investigate a specific family of \wqos built inductively as
described in \Cref{def:fam:elementary-wqo} that we call \reintro{elementary
\wqos}. While the ordinal invariants of $\Pf(A)$ remain non-functional with
respect to those of $A$ for \kl{elementary \wqos}, we showed that by introducing
\emph{weakened} versions of the usual ordinal invariants (see
\Cref{def:approximate-invariants}), one could recover compositional
computability of weakened and non-weakened invariants over elementary \wqos, as
shown in \Cref{sec:algebra-well-behaved}. Namely, we provide in
\Cref{tab:fam:h-pf-computation,tab:computing-elementary} (respectively in
page~\pageref{tab:fam:h-pf-computation} and
page~\pageref{tab:computing-elementary}) recursive equations to compute the
ordinal invariants of elementary \wqos.

\begin{table}[ht]
	\centering
    \caption{Tight upper and lower bounds for the ordinal invariants of
    $\Pf(A)$.}
	\begin{tabular}{c c c l}
		\toprule
		\textbf{Invariant} & \textbf{Lower Bound} & \textbf{Upper Bound} & %
                \\
		\midrule
		 $\o(\Pf(A))$       & $1 + \o(A)$      & $2^{\o(A)}$ &
		Thm.\&Lem.~\ref{lem:pf-plus-sqcup}.(3), \ref{thm:pf-mot}, \ref{thm:pf-o-ub-tight}
		\\
		$\h(\Pf(A))$       & $1 + \h(A)$       & $2^{\h(A)}\cdot\omega$ &
		Thm.\&Lem.~\ref{lem:pf-plus-sqcup}.(3),
                \ref{thm:pf-height}, \ref{lem:h-FSim}, \ref{lem:h-FSim-m}
		\\
		$\w(\Pf(A))$\footnotemark & $2^{\w(A)}$         & none                 &
		Exmp.\&Thm.~\ref{exmp:w-low-2}, \ref{thm:pf-width}, \ref{thm:pf-w-lb-tight}
                \\
		\bottomrule
	\end{tabular}
	\label{table:ordinal-invariants-pf}
\end{table}
\footnotetext{Assuming $\omega\leq\w(A)$.}

\paragraph*{Outline of the article.} \Cref{sec:basics} recalls the definitions
that will be used in this article. \Cref{sec:up-lower-bounds} establishes upper and
lower bounds for the \kl{ordinal invariants} on $\Pf(A)$. \Cref{sec:tightness}
provides families of \wqos attaining those bounds. Finally,
\Cref{sec:algebra-well-behaved} shows that restricting ourselves to
\kl{elementary \wqos}, we can compute the \kl{ordinal invariants} and their
weakened versions compositionally.

\paragraph*{Genesis of this article.} This article grew from unpublished notes
(2013) by Abriola, Schmitz and Schnoebelen who investigated the \kl{m.o.t.} of
$\Pf(A)$, established the upper and lower bounds we give in \Cref{ssec-mot-Pf},
and proved their tightness. The observation that $\o(\Pf(A))$ is not a function
of $\o(A)$ spurred research in the width invariant, with the hope that
$\o(\Pf(A))$ could be better characterised when both $\o(A)$ and $\w(A)$ are
known but this idea proved inconclusive. Then, in 2020, Halfon and Lopez revived
this line of work and gave bounds for the height and width of $\Pf(A)$,
together with proofs of tightness. In 2022, Vialard joined the group and showed
how one can determine exactly the values of all three invariants for the rather
large family of \wqos given in \Cref{sec:algebra-well-behaved}. It was then
decided by all six contributors that this material will profit from being collected in a
single text.

\section{Basics}
\label{sec:basics}

We assume that the reader is familiar with basic notions on ordinals and quasi-orders.
We recall here the definitions and notations that will be used throughout the
article.

\paragraph*{Well-quasi-orderings.} \AP A sequence $(x_i)_{i \in \mathbb{N}}$ is
\intro{good} in a quasi-order $(A,\leq)$ when there exists an increasing pair
$x_i \leq x_j$ with $i < j$. Otherwise it is \intro{bad}. When the ordering can
be inferred from the context, we may write just $A$ instead of $(A,\leq)$ for a
quasi-order. A \intro{well-quasi-order} (or \wqo) is a quasi-order in which
there are no infinite \kl{bad sequences}.  A \wqo can alternatively be defined
as a quasi-order which is \intro{well-founded} (it has no infinite decreasing
sequences) and satisfies the \intro{finite antichain condition} (it has no
infinite \intro{antichains}: sets of pairwise incomparable
elements).\footnote{We use $a\bot b$ to denote that $a$ and $b$ are
incomparable, i.e., that $a\not\leq b \land b\not\leq a$.}

\AP If a \wqo $(A, \leq)$ is such that $\leq$ is anti-symmetric, then it is a
\intro{well-partial-order} (\wpo). Any \wqo can be turned into a \wpo by
quotienting by the equivalence relation ${\leq}\cap{\geq}$, also
denoted with $\equiv$. We say that two
\wqos are \intro{{\wpo}-isomorphic} (denoted with $\intro*\wpoiso$) when the
\wpos obtained through quotienting are isomorphic (denoted with $\cong$). We prefer working with \wqos
because constructions like taking subsets or multisets yield \wqos even when applied
to \wpos.

\AP When $\alpha$ is an ordinal, we may use  $\alpha$  to denote the associated
well-order $(\alpha,\in)$, which is a \wqo. For $k\in\Nat$ we let
$\intro*\FAnt{k}$ denote a \wqo of $k$ incomparable elements, i.e., a size-$k$
\kl{antichain}.

\paragraph*{Classical operations on wqos.} %
We will now recall the definitions
of the operations on \wqos that are used in this article; for a more thorough
survey on these constructions and their ordinal invariants, we redirect the
reader to the work of \cite{dzamonja2020}. For the rest of this section, we
will assume that $(A,\leq_A)$ and $(B,\leq_B)$ are two \wqos.

\AP The \intro{disjoint sum} $(A \intro*\dunion B,\leq_\sqcup)$ and the
\intro{lexicographic sum} $(A \intro*\lexsum B,\leq_+)$ both have for support
the disjoint union of $A$ and $B$.  The ordering of $A\dunion B$ is defined as
$\leq_A\cup\leq_B$, whereas the ordering of $A \lexsum B$ is
${{\leq_A}\cup{\leq_B}\cup\setof{ (a, b) }{a\in A, b\in B}}$.

\begin{example}
  \label{ex:disjoint-union}
  Consider the \wqos $\FAnt{2}$ and $\FAnt{3}$ made of 2 and 3
  incomparable elements, respectively. Their disjoint sum
  $\FAnt{2} \sqcup \FAnt{3}$ is a \wqo made of 5 incomparable
  elements, i.e., $\FAnt{5}$. Their lexicographic sum
  $\FAnt{2} \lexsum \FAnt{3}$ is a \wqo made of 2 levels: the
  bottom level contains 2 incomparable elements, and the top
  level contains 3 incomparable elements, each greater than
  elements of the bottom level.
\end{example}

It is easy to see that the \kl{disjoint sum} $\Dunion_{1 \leq i \leq n} A_i$ of
finitely many \wqos is a \wqo, and it is immediate that an infinite disjoint
sum (of non-empty) \wqos is not a \wqo. For the \kl{lexicographic sum}, the
infinite case is more interesting:  given an ordinal $\alpha$ and an
$\alpha$-indexed family of \wqos $(A_\gamma)_{\gamma<\alpha}$, their
\kl{lexicographic sum} $\Lexsum_{\gamma<\alpha} A_\gamma$ remains a \wqo.

\AP
The \intro{Cartesian product} $(A \intro*\cartprod B,\leq_\times)$ and the \intro{lexicographic product}
$(A \intro*\lexprod B,\leq_{\cdot})$ both have for support
$A \times B$, and their orderings are obtained as follows:
\begin{align*}
    (a,b) \leq_\times (a',b') &\equivdef a\leq_A a' \wedge b \leq_B b' \:,
\\
    (a,b)\leq_\cdot (a',b') &\equivdef b \leq_B b' \wedge (b\equiv_B b' \implies a\leq_A a') \: .
\end{align*}
In this document, we will use $A^{\times n}$ to denote the
\kl{Cartesian product} of $n$
copies of $A$, with $n$ finite.

\begin{example}
  \label{ex:cartesian-product}
  Consider the \wqos $A_3$ made of 3 elements $a_1<a_2<a_3$ and
  $\FAnt{2}$ made of 2 incomparable elements.
  Their Cartesian product $\FAnt{2} \cartprod A_3$ is a \wqo
  made of 6 elements: two chains $a_1<a_2<a_3$ side by side.
  Their lexicographic product $\FAnt{2} \lexprod A_3$
  is a \wqo made of 3 levels: each level containing a copy of $\FAnt{2}$,
  and each element of a level being greater than all elements of the
  lower levels.
  A graphical representation of these two products is given in
  \Cref{fig:cartesian-lexicographic-product}.
\end{example}

\begin{figure}
  \centering
  \begin{tikzpicture}
    \begin{scope}[xshift=-3cm]
      \foreach \x in {1,2} {
        \foreach \y in {1,2,3} {
          \node[inner sep=1pt,fill=black,circle,draw] (a\x\y) at (\x, \y) {};
        }
      }
      \foreach \x in {1,2} {
        \foreach \y in {1,2} {
          \draw (a\x\y) -- (a\x\the\numexpr\y+1\relax);
        }
      }
    \end{scope}
    \begin{scope}[xshift=3cm]
      \foreach \y in {1,2,3} {
        \foreach \x in {1,2} {
          \node[inner sep=1pt,fill=black,circle,draw] (b\x\y) at (\x, \y) {};
        }
      }
      \foreach \y in {1,2} {
        \draw (b1\y) -- (b1\the\numexpr\y+1\relax);
        \draw (b2\y) -- (b2\the\numexpr\y+1\relax);
        \draw (b1\y) -- (b2\the\numexpr\y+1\relax);
        \draw (b2\y) -- (b1\the\numexpr\y+1\relax);
      }
    \end{scope}
  \end{tikzpicture}
  \caption{\kl{Cartesian product} (left) and \kl{lexicographic product} (right)
    of the \wqos $A_3$ and $\FAnt{2}$ from \Cref{ex:cartesian-product}.
    The representation is a Hasse diagram where edges are oriented
    upwards.}
  \label{fig:cartesian-lexicographic-product}
\end{figure}

It is known that both products are \wqos when $A$ and $B$ are. We would like to
emphasize the fact that our definition of the \kl{lexicographic product} gives
priority to the second component of pairs, contrary to the convention for words
in a dictionary: this choice aligns \kl{lexicographic product} with the usual
product of ordinals.

\AP
The set of finite sequences (or \intro{words}) over $A$, written $A^\seq$, is ordered
using the \intro{word embedding relation} defined via:
\begin{equation*}
    \bm{u} = u_1\cdots u_n \leq_\seq \bm{v} = v_1\cdots v_{m}
    \equivdef
    \exists 1\leq f_1<f_2<\cdots<f_n\leq m \text{ such that }
    \forall 1 \leq i \leq n, u_i \leq_A v_{f_i}
    \: .
\end{equation*}
In the above definition, $n$ and $m$ are the lengths of $\bm{u}$ and $\bm{v}$,
respectively, and $f$ is an \kl(word){embedding} of $\bm{u}$ into $\bm{v}$.
We use $\epsilon$ to denote the empty word, of length $0$.

\begin{remark}
  \label{rem:word-embedding-inductive}
  The \kl{word embedding relation} can be defined inductively as follows:
  $\epsilon \leq_\seq \bm{v}$ for any word $\bm{v}$, and
  $a \bm{u} \leq_\seq b \bm{v}$ iff either $a \leq_A b$ and
  $\bm{u} \leq_\seq \bm{v}$, or $a \bm{u} \leq_\seq \bm{v}$.
\end{remark}

\AP The \intro{set of finite multisets} $(\intro*\Mul(A),\leq_\diamond)$ of
elements in $A$ ordered with \intro{multiset embedding}, corresponds to
$(A^\seq,\leq_\seq)$ quotiented by the following equivalence relation:
$\bm{u}\equiv \bm{v}$ iff $\bm{v}$ can be obtained from $\bm{u}$ by permuting
the order of the letters. Equivalently $\bm{u}\leq_\diamond \bm{v}$ iff there
exist some permutations $\bm{u}\equiv \bm{u'}$ and $\bm{v}\equiv \bm{v'}$ with
$\bm{u'}\leq_\seq \bm{v'}$ (actually, a single permutation is always
sufficient).

\begin{remark}
  \label{rem:multiset-not-mana-dershowitz}
  The \kl{multiset embedding} relation, as defined above,
  does \emph{not} coincide with another classical ordering
  used on multisets due to Manna and Dershowitz~\citep{dershowitz79}.
\end{remark}

\begin{center}
  \textasteriskcentered
  \textasteriskcentered
  \textasteriskcentered
\end{center}

\AP
Let us now introduce the main operator of this article, that is the
\intro{finitary powerset construction}.  We write $\intro*\Pf(A)$ for the collection
of finite subsets\footnote{This collection is more usually denoted
$\mathcal{P}_{\text{fin}}(A)$ in the literature, or also $\text{Fin}(A)$,
$[A]^{<\omega}$, and $\mathcal{P}_{<\omega}(A)$ in some subfields.}
of $A$, with typical elements $S,S',\ldots$ and
endow it with the \intro{Hoare embedding} relation (also known as the
\reintro{domination quasi-ordering}, or the \reintro{lower preorder}), defined via
\begin{equation}
  \label{eq:hoare-ordering}
  \tag{Hoare}
S \intro*\hoareleq S' \;\equivdef\; \forall a \in
S \: \exists b\in S' \: a\leq_A b \:.
\end{equation}
It is well-known that $(\Pf(A),\hoareleq)$ is a \wqo when $A$ is whereas, as
shown by \citet{rado54}, the full powerset $\intro*\Pfull(A)$ (that considers infinite subsets as
well) needs not be.

\begin{remark}
The literature  considers other ways of ordering the powerset of ordered
sets, see, e.g., \citep{marcone2001}. There exist the \emph{Smyth ordering}, or
\emph{upper preorder}, defined
with $S \intro*\smythleq S' \;\equivdef\; \forall b \in S' \: \exists a\in S
\: a\leq_A b$, and an Egli-Milner ordering obtained as the intersection
of $\hoareleq$ and $\smythleq$.

There is a duality between $\hoareleq$ and $\smythleq$: for
$S_1,S_2\subseteq A$ one has $S_1\hoareleq S_2 \iff A\setminus(\dc
S_1)\smythleq A\setminus (\dc S_2)$, where $\dc S$ denotes the
\kl{downward-closure} of a subset of $A$. In this article we neither consider
the Smyth nor Egli-Milner orderings since they do not in general
give rise to a well-quasi-ordered
$\Pf(A)$.
\end{remark}

\paragraph*{About ordinal arithmetic.} To measure the complexity of \wqos, we
will use ``ordinal invariants,'' and therefore need to manipulate ordinals and
their arithmetic. We suppose well-known the operations of sum $\alpha+\beta$,
product $\alpha\cdot\beta$, natural sum $\alpha\oplus\beta$, and natural
product $\alpha\otimes\beta$ on ordinals, and refer to, e.g., \cite{fraisse86} for a
more complete presentation of ordinal arithmetic. Let us point out particular
ordinals that will often be special cases in our results and proofs.

\begin{definition}%
    An ordinal $\alpha$ is
    \begin{itemize}
        \item an \intro{$\varepsilon$-number}
        whenever $\om^{\alpha} = \alpha$;
        \item \intro{additively indecomposable} when
        for any $\beta < \alpha$
        and $\gamma < \alpha$, $\beta\oplus\gamma<\alpha$.
        Alternatively, \kl{additively indecomposable} ordinals are of the form
        $\om^{\alpha'}$ with $\alpha'$ any ordinal.
        \item \intro{multiplicatively indecomposable} when
        for any $\beta < \alpha$
        and $\gamma < \alpha$, $\beta\otimes\gamma<\alpha$.
        Alternatively, \kl{multiplicatively indecomposable} ordinals (besides $2$) are of the form $\om^{\alpha'}$ with $\alpha'$ \kl{additively indecomposable}.
    \end{itemize}
\end{definition}

As an abuse of language, we will simply write ``\reintro{indecomposable}'' for
``\kl{multiplicatively indecomposable}'' in the rest of this article, and always
rule out 0, 1 and 2 from our notions of \kl{indecomposable} ordinals in order to
simplify our statements, as these will often be treated separately.

Since some of our results rely heavily
on $2$-exponentiation (see \Cref{table:ordinal-invariants-pf}),
let us briefly recall that
ordinal exponentiation is defined via $\alpha^0 \defined 1$,
$\alpha^{\beta+1} \defined \alpha^\beta\cdot\alpha$ and, for $\lambda$ a limit ordinal,
$\alpha^\lambda \defined \sup_{\gamma<\lambda}\alpha^\gamma$.  Exponentiation satisfies
both $\alpha^{(\beta+\gamma)}=\alpha^\beta\cdot\alpha^\gamma$ and
$(\alpha^\beta)^\gamma=\alpha^{(\beta\cdot\gamma)}$, which entails the following useful fact:

\begin{fact}[$2$-exponentiation]
    \label{fact:two-exponentiation}
    If $n<\om$ and $\alpha = \om \cdot \alpha' + n$ then $2^{\alpha} = \om^{\alpha'}\cdot 2^n$.
\end{fact}
Now since any ordinal can be written (in a unique way) as $\alpha = \om\cdot\alpha'+n$ with
$n<\om$, \Cref{fact:two-exponentiation} implies that $2^\alpha$ is \kl{additively indecomposable}
if, and only if, $\alpha$ is a limit ordinal.

\AP
When constructing examples of \wqos with specific ordinal invariants, we will often
rely on the \intro{Cantor Normal Form} (CNF) of ordinals. Any ordinal $\alpha$ can be
written uniquely as a finite sum
\begin{equation*}
    \alpha = \sum_{i< n} \om^{\alpha_i}
\end{equation*}
with $\alpha_0 \geq \alpha_1 \geq \cdots \geq \alpha_{n-1}$ (and $n=0$
for $\alpha=0$). The
ordinals $\alpha_i$ are the exponents of the CNF of $\alpha$.
For an \kl{$\varepsilon$-number} $\alpha$, its \kl{CNF} has a single term, i.e., $\alpha = \om^{\alpha}$.

\paragraph*{Ordinal invariants.} \AP We follow the path of
\citet[\S~4]{kriz90b} to introduce \intro{ordinal invariants}. For any \wqo\
$(A,\leq_A)$, we define $\intro*\BadSeq(A)$ (respectively $\intro*\DecSeq(A)$
and $\intro*\IncSeq(A)$) as the tree of bad sequences (respectively strictly
decreasing sequences and antichain sequences) of $A$ ordered by inverse prefix
order: in each tree the empty sequence is the root, and if a sequence $s$ is a
prefix of a sequence $t$, then $s \geq t$.

Observe that, since $A$ is a \wqo (hence satisfies the \kl{finite antichain
condition} and is \kl{well-founded}),  $\BadSeq(A)$ is a tree without infinite
branches, i.e., is well-founded, and so are $\DecSeq(A)$ and $\IncSeq(A)$ since
they are subtrees of $\BadSeq(A)$. However, these trees may be
infinite by way of having infinite branching.

\AP As is standard with well-founded trees, an \intro{ordinal rank}
$\intro*\orank(s)$, defined from bottom to top, is associated with any node $s$ of $T$.  If $s$ is a
leaf, then $\reintro*\orank(s)$ is defined as $0$. Otherwise,
$\reintro*\orank(s)$ is defined as the supremum of the ranks of its successors
plus one, i.e., $\sup\setof{\reintro*\orank(t) + 1}{t \leq s \text{ in } T}$.
Since $T$ can be infinitely branching, $\orank(s)$ may be an infinite ordinal.
The \reintro{rank of $T$} is defined as the \kl{rank} of its root.

\begin{definition}[Ordinal invariants]
    The \intro{maximal order type} (m.o.t.),
    \intro{height} and \intro{width}
    of a \wqo $A$ are the \kl{ranks} of $\BadSeq(A)$, $\DecSeq(A)$ and
    $\IncSeq(A)$,  respectively. They are denoted $\intro*\o(A)$, $\intro*\h(A)$ and $\intro*\w(A)$.
\end{definition}

\begin{remark}
  Since $\DecSeq(A)$ is \kl{well-founded} when $A$ is \kl{well-founded},
  the definition of \kl{height} can be extended to \kl{well-founded
  quasi-orders}.
Similarly, the definition of \kl{width} applies to any quasi-order
satisfying the \kl{finite antichain condition}~\citep{dzamonja2020}.
\end{remark}

\AP In accordance with the inductive definition of the \kl{ordinal rank}, the
\kl{ordinal invariants} can be computed using so-called \intro{descent
equations}. For a relation symbol $\someRelation$ among, e.g.,
$<,\leq,\not\geq,\bot,\ldots$, and some element $x$
of a \wqo $A$, we define the \intro{residual}
$A_{\someRelation x}=\setof{y \in A}{y \someRelation x}$. Then, with the above notations the following
inductive characterisations can be derived:
\begin{align}
    \label{eq-Des-o}\tag{Do}\o(A)&=\osup{ \o(A_{\not\geq x})+1 }{x\in A} \:,\\
    \label{eq-Des-h}\tag{Dh}\h(A)&=\osup{ \h(A_{< x})+1 }{x\in A} \:,\\
    \label{eq-Des-w}\tag{Dw}\w(A)&=\osup{ \w(A_{\perp x})+1 }{x\in A} \:.
\end{align}
The notion of \kl{residual}  extends to subsets $B\subseteq A$
by considering the intersection of the \kl{residuals} $A_{\someRelation y}$ where $y$
ranges over $B$, namely:
\begin{equation*}
    A_{\someRelation B} \eqdef \bigcap_{y\in B} A_{\someRelation y} \:.
\end{equation*}

\AP The reader may notice that $A_{\leq x}$  is in fact the \intro{downward closure} of
$\{x\}$ in $A$, a basic and useful notion that we denote with
$\intro*\dwc_A \set{x}$ or $\dwc_A x$ as is common.  However, residuation and closure are
different concepts: the downward closure of some subset $B
\subseteq A$, written $\dwc_A B$, is the \emph{union} of the closures $\dwc_A \set{y}$
for $y \in B$, hence is \emph{not} the same as $A_{\leq B}$ (unless $B$ is a singleton).
We write more simply $\dwc x$ and $\dwc B$ when the \wqo $A$ is clear from
the context.  Note that $\dwc_A A = A$ (closures are extensive), and that $\dwc
\emptyset = \emptyset$ (they preserve the empty set).
A set $B \subseteq A$ is \intro{downwards closed} when $\dwc_A B = B$.

Using the \kl{descent equations}, we can compute our first ordinal invariants
and relate them to each other. For instance, the ordinal invariants of the
finite partial-order of \cref{fig:hasse-width-height-size} are exactly its
size, height and width, i.e., $\o(A)=6$, $\h(A)=4$ and $\w(A)=3$. Similarly, it
is relatively straightforward to prove that for $A = \om$ (the natural numbers
with the usual ordering), then $\o(A) = \h(A) = \om$, and $\w(A) = 1$.

One important fact about the \kl{ordinal invariants} that
is worth mentioning is that they only count ``non-equivalent'' elements
(strictly below, non-comparable, or not above), hence are oblivious
to quotienting by equivalence. Hence, working with a \wqo or with
its associated \wpo does not change the values of the invariants:
\begin{fact}
  \label{fact:invariants-wpo-isom}
  Let $A$ and $B$ be two \wqos. If $A \wpoiso B$ then
  $\someInvariant(A) = \someInvariant(B)$
  for any invariant $\someInvariant$ among $\{\o, \h, \w\}$.
\end{fact}

We summarize the current knowledge on ordinal invariants in
\Cref{tab:fam:la-famille}.\footnote{The table uses some ad-hoc
ordinal operations, like $\alpha^{\pm}$ or $\alpha\hplus\beta$, for
conciseness and clarity: they all appear in earlier papers (but
sometimes the notation is new) and their definition is recalled
in \Cref{fig:fam:ordinal-operations} (page~\pageref{fig:fam:ordinal-operations}).}
As one can see, for most operations on \wqos, the ordinal invariants can
be expressed as a function of the invariants of the arguments. Actually,
this is even better, one can compute  any specific invariant
from the same invariant of the arguments.

\begin{table}[!hb]
    \centering
	\caption{How to compute ordinal invariants compositionally \citep{dzamonja2020}.}
		\renewcommand{\arraystretch}{1.5}
		\begin{tabular}{c c c c}
			\toprule
			\textbf{Space} & \textbf{M.O.T.}            & \textbf{Height}      &
			\textbf{Width}                                                                            \\
			\midrule
		$\alpha$       & $\alpha$                   & $\alpha$             & $1$                  \\
			$A\dunion B$    & $\o(A)\oplus\o(B)$         & $\max (\h(A),\h(B))$ & $\w(A)\oplus\w(B)$   \\
			$A \lexsum B$        & $\o(A)+\o(B)$              & $\h(A)+\h(B)$        & $\max (\w(A),\w(B))$ \\
			$A\cartprod B$    & $\o(A)\otimes\o(B)$        & $\h(A)\hplus \h(B)$  & \emph{Not functional}             \\
			$A\lexprod B$     & $\o(A)\cdot\o(B)$~\footnotemark & $\h(A)\cdot \h(B)$ & $\w(A)\odot\w(B)$    \\

			\addlinespace
			$\Mul(A)$
			               & $\om^{\widehat{\o(A)}}$
			               & $h^\star(A)$
			               & $\o(\Mul(A))$~\footnotemark
			\\
			\addlinespace
			$A^\seq$
			               & $\om^{\om^{(\o(A)^\pm) }}$
			               & $h^\star(A)$
			               & $\o(A^\seq)$~\footnotemark
			\\
			\addlinespace
			$\Pf(A)$       & \unknown                   & \unknown             & \unknown
			\\
			\bottomrule
		\end{tabular}
	\label{tab:fam:la-famille}
\end{table}
\addtocounter{footnote}{-2}
\footnotetext{Assuming that $\o(B)$ is a limit. See~\cite{vialard-oPcdotQ} for the general case.}
\addtocounter{footnote}{1}
\footnotetext{Assuming that $\o(A)=\omega^\alpha$ is \kl{additively indecomposable} $\geq\omega$.}
\addtocounter{footnote}{1}
\footnotetext{Assuming that $\o(A)>1$.}

\begin{example}
  \label{ex:computing-compositionally}
	Consider
	$A = (\om \lexsum \om) \;\dunion\; (\om \lexsum \om)$.
  Using \Cref{tab:fam:la-famille}, we can compute its
  ordinal invariants compositionally as follows:
  $\o(A) = \o(\om + \om) \oplus \o(\om + \om)
  = \om \cdot 4$.
  Similarly, $\h(A) = \om \cdot 2$,
  and $\w(A) = 2$.
\end{example}

Unfortunately, both the \kl{Cartesian product} and the \kl{finitary powerset}
do not belong to this nice picture, as witnessed by the ``Not functional'' and
``Unknown'' entries in \Cref{tab:fam:la-famille}.

\begin{figure}[ht]
\begin{align*}
\alpha \hplus \beta &\defined \sup \Setof{\alpha'\oplus\beta'}{\alpha'<\alpha,\beta'<\beta}\:,
\\
\alpha^\pm  &\defined
 		\begin{cases}
 			\alpha - 1 & \text{ if $\alpha$ is finite,} \\
 			\alpha + 1 & \text{ if $\alpha = \varepsilon + n $ with $\varepsilon$ an $\varepsilon$-number and $n<\omega$,} \\
 			\alpha     & \text{ otherwise,}
 		\end{cases}
\\
h^\star(A) &\defined
 		\begin{cases}
 			\h(A) & \text{ if } \h(A) \text{ is additively
                        \kl{indecomposable} } \geq\om \:, \\
 			\h(A)\cdot\om           & \text{ otherwise,}
 		\end{cases}
\\
\oprim{\alpha} &\defined \begin{cases}
 			\alpha + 1 & \text{ if } \alpha = \varepsilon
                        + n  \:, \\
 			\alpha     & \text{ otherwise.}
 		\end{cases}
\\
\widehat{\alpha} &\defined \omprim{\alpha_1}+ \dots +
\omprim{\alpha_n} \text{ when }
 		\alpha = \om^{\alpha_1} + \dots + \om^{\alpha_n} \:,
\\
\alpha\hprod\beta &\text{ defined via }
 	\begin{cases}
 		\alpha \hprod 0 \defined 0 \:,  \\
 		\alpha \hprod (\beta+1) \defined (\alpha \hprod \beta)
                 \nadd \alpha \:, \\
		\alpha \hprod \lambda \defined \sup_{\gamma < \lambda}
                (\alpha \hprod \gamma + 1)\text{ for $\lambda$ limit.}
        \end{cases}
\end{align*}
	\caption{Definition of the notations used in \Cref{tab:fam:la-famille}.}
	\label{fig:fam:ordinal-operations}
\end{figure}

\paragraph*{Techniques for ordinal invariants.}
When compositional techniques fail, one can resort to more general methods
to compute or estimate the ordinal invariants of \wqos. In the remainder of
this section, we introduce
the basic toolbox  used in this article.
First,
since $\DecSeq(A)$ and $\IncSeq(A)$ are subtrees of $\BadSeq(A)$,
we always have the following inequalities:
\begin{equation}
  \label{eq:basic-inequalities}
  \h(A) \leq \o(A) \quad\text{and}\quad \w(A) \leq \o(A) \:.
\end{equation}

In the finite case, one also knows that $\o(A) \leq \h(A) \times \w(A)$, where
$\times$ is the usual multiplication of natural numbers. As shown
by \citeauthor*{kriz90b}, this  generalizes to
  the infinite  case, using the natural product, $\otimes$, on ordinals:

\begin{therm}[{\protect\citet[Thm.~4.13]{kriz90b}}]
	\label{thm:kt}
	For any \wqo $A$,
		 $\o(A) \leq \h(A) \otimes \w(A)$.
\end{therm}

\Cref{thm:kt} is often used when $\o(A)$ is multiplicatively
indecomposable in view of the following corollary:

\begin{corollary}
\label{coro:kt}
Assume that $\o(A)$ is \kl{multiplicatively indecomposable}.
If $\h(A)<\o(A)$ then $\w(A)=\o(A)$, and if $\w(A)<\o(A)$ then $\h(A)=\o(A)$.
\end{corollary}

Another useful technique to establish inequalities between ordinal invariants
is to compare the invariants of different \wqos. The most straightforward way
to relate two \wqos for us will be \kl{reflections} and their special cases,
\kl{substructures} and \kl{augmentations}.

\begin{definition}[Augmentation]
	\label{def:augmentation}
	A quasi-order $(A,\leq_A)$ is an \intro{augmentation} of $(B,\leq_B)$
	whenever $A = B$ and ${\leq_B} \subseteq {\leq_A}$.
	We write this relation as $B \intro*\leqaug A$.
\end{definition}

\begin{definition}[Substructure]
	A quasi-order $(A, \leq_A)$ is a \intro{substructure} of a quasi-order $(B, \leq_B)$
	whenever $A \subseteq B$ and $\leq_A$ is the restriction
        $\leq_B\cap(A\times A)$ of $\leq_B$
	to $A$. This relation is written $A \intro*\leqstruct B$.
\end{definition}

In the remainder of this article we
often write statements like, e.g., ``$A$ is a \kl{substructure} of  $A\cartprod
A$'', when we really mean ``$A$ is \emph{isomorphic to a} \kl{substructure} of
$A\cartprod A$'', or even \kl{\wpo-isomorphic} to a \kl{substructure} of
$A\cartprod A$, since these do not influence the ordinal invariants (recall
\cref{fact:invariants-wpo-isom}).

\begin{definition}[Reflection]
	A mapping
	$f \colon A \rightarrow B$ between two \wqos
	is a \intro{reflection} if $f(x)\leq_B f(y)$ implies $x \leq_A y$.
	We write $A \intro*\refl B$ when there exists a reflection from $A$ to $B$.
\end{definition}

\begin{example}
  \label{ex:reflection}
  Let $A$ be any \wqo. The mapping $f: A \to \Pf(A)$ defined by
  $f(x) = \{x\}$ is a \kl{reflection} from $A$ to $\Pf(A)$ since
  $f(x) \hoareleq f(y)$, i.e., $\{x\} \hoareleq \{y\}$,
  implies $x \leq_A y$.
\end{example}

In general the \kl{ordinal invariants} behave monotonically with
reflections and their special cases, as summarized in
\Cref{lem:invariants-monotonicity}.
\begin{lemma}
\label{lem:invariants-monotonicity}
For any \wqos $A,B$,\begin{itemize}
\item if $A\leqstruct B$ then $\someInvariant (A)\leq \someInvariant(B)$ for $\someInvariant  \in\set{\o,\w,\h}$.
\item if $A\geqaug B$ then $\someInvariant(A)\leq \someInvariant(B)$ for $\someInvariant\in\set{\o,\w}$.
Moreover, if $A$ is a \wpo then $\h(A)\geq \h(B)$.
\item if $A\refl B$ then $\someInvariant(A)\leq \someInvariant(B)$ for $\someInvariant\in\set{\o,\w}$.
\end{itemize}
\end{lemma}
The exception with $\h$ is that, when $A$ is an augmentation of $B$,
the pairs $(a,b)$ in $\leq_A\setminus\leq_B$ may give rise to
new descending sequences,  and potentially to a larger height,  but
they may also give rise to new pairs of equivalent elements (if $b<_B a$
then $a\equiv_A b$) with the potential of reducing some strictly
decreasing sequences from $A$, leading to a diminished height.

As many proofs will rely on the \kl{descent equations}
(\ref{eq-Des-o}--\ref{eq-Des-w}), it will be particularly useful for us
to be able to find \kl{substructures} with prescribed ordinal invariants when
proving lower bounds.
This is always possible, as stated in \Cref{lem:desc-eq}.
\begin{lemma}
\label{lem:desc-eq}
For any ordinals $\beta<\alpha$, any $\someInvariant\in\set{\o,\w,\h}$, and any \wqo $A$ such that $\someInvariant(A)=\alpha$, there exists $B\leqstruct A$ such that $\someInvariant(B)=\beta$.
\end{lemma}
\begin{proof}
It is well-known (see, e.g., \citet{wolk67}) that when a well-founded
tree has rank $\alpha$ then
every $\beta<\alpha$ is the rank of some node of the tree. Thus if
$\beta<\o(A)$, the tree of bad sequences has a node $s$ (a bad sequence) of rank
$\beta$. If we write $S$ for the set of elements listed in $s$, the subtree
rooted at $s$ is isomorphic to $\Bad(A_{\not\geq S})$ so
$\beta=\o(A_{\not\geq S})$ and we conclude by noting that the residual
$A_{\not\geq S}$ is a substructure of $A$. Finally, the same reasoning
applies to $\h(A)$ and $\w(A)$ using the trees $\Dec(A)$ and $\Inc(A)$
and residuals of the form $A_{<S}$ and $A_{\perp S}$.
\end{proof}

Notice that $\Pf(A)$ is a quotient of $\Mul(A)$ through the map $ M\in\Mul(A)
\mapsto \setof{ x \in A }{ M(x) \geq 1 }$. Unfortunately quotients do not
preserve the ordinal invariants, and $\Mul(A)$ and $\Pf(A)$ actually exhibit
strikingly different behaviors.

\paragraph*{Finitary powerset.} Let us finish this preliminary section by
collecting few basic facts about the \kl{finitary powerset} constructions that
can be immediately derived from our above tools and definitions. These are
folklore results, but will be used extensively in the remainder of this article.

Like all the previously recalled operations on \wqos, the \kl{finitary
powerset} construction is monotonic with respect to the orderings associated
with \kl{augmentations}, \kl{substructures} and \kl{reflections}: if
$A\leqaug B$ then $\Pf(A)\leqaug \Pf(B)$ etc.

For some proofs, it will be easier to work with \wpos rather than \wqos.
Unfortunately, the \kl{finitary powerset} construction does not preserve \wpos
and can introduce equivalence classes. However $\Pf(A)$ is
\kl{{\wpo}-isomorphic} to $\intro*\Ant(A)$, the set of (finite) antichains
ordered with the \kl{Hoare embedding}, through the map associating with any
finite $S\in\Pf(A)$ the set $\max S$ of its maximal elements. If $A$ is a \wpo
then $\Ant(A)$ is a \wpo too. Similarly, $\Pf(A)$ is wpo-isomorphic to $D(A)$,
the set of \emph{finitely generated lower segments} of $A$, ordered by
inclusion \citep{altman2025}.

We list in \Cref{lem:pf-plus-sqcup} some simple isomorphisms (or
\kl{{\wpo}-isomorphisms}) involving the \kl{finitary powerset} construction
together with \kl{disjoint unions}, \kl{lexicographic sums}, and ordinals.

\begin{lemma}
\label{lem:pf-plus-sqcup}
Let $A,B$ be \wqos, and $\alpha$ be an ordinal.
Then,
  \begin{enumerate}
    \item
      $\Pf(A \lexsum B) \wpoiso \Pf(A) + (\Pf(B) \setminus \set{\emptyset})$,
    \item $\Pf(A \dunion B) \cong \Pf(A) \cartprod \Pf(B)$, and
    \item $\Pf(\alpha) \wpoiso 1 + \alpha$.
  \end{enumerate}
\end{lemma}
\begin{proof}
Any finite subset $S$ of $A \lexsum B$ can be decomposed into a subset $S_A$ of
$A$ and a subset $S_B$ of $B$. For any two subsets $S,S'$, if
$S'_B\neq\emptyset$, then $S\hoareleq S'$ iff $S_B \hoareleq S'_B$.
Otherwise, when $S'_B=\emptyset$, then $S\hoareleq S'$ iff $S_B=\emptyset$
too and $S_A\hoareleq S_A'$. Thus the function mapping $S$ to  $S_B$
if $S_B\neq\emptyset$, and to $S_A$ otherwise, is a monotonic \kl{reflection}
from $\Pf(A\lexsum B)$ to $\Pf(A) \lexsum (\Pf(B) \setminus \set{\emptyset})$.

Similarly, any finite subset $S$ of $A\dunion B$ can be decomposed into a
  subset $S_A$ of $A$ and a subset $S_B$ of $B$, such that $S\hoareleq S'$ iff
  $S_A\hoareleq S'_A$ and $S_B\hoareleq S'_B$ for any finite subset $S,S'$ of
  $A\dunion  B$. Thus the function mapping $S$ to the pair $(S_A,S_B)$
  is an isomorphism between $\Pf(A\dunion B)$ and $\Pf(A)\cartprod \Pf(B)$.

Finally, we observed earlier that $\Pf(\alpha)$ is \kl{{\wpo}-isomorphic} to
$\Ant(\alpha)$. However, any non-empty finite subset of $\alpha$ has a unique
maximum, hence, every (non-empty) antichain of $\alpha$ is a singleton set.
When also the empty set is counted we obtain $\Ant(\alpha) \cong 1 + \alpha$.
\end{proof}

\section{Upper and Lower Bounds for $\Pf(A)$}
\label{sec:up-lower-bounds}

Our first claim is that the ordinal invariants of $\Pf(A)$ cannot be expressed as a function of $\o(A)$, $\h(A)$, and $\w(A)$, as
witnessed in \Cref{ex:non-functional}.

\begin{example}
	\label{ex:non-functional}
	Consider
	$A_1 = (\om + \om) \;\sqcup\; (\om + \om)$
	and
	$A_2 = (\om \sqcup \om)\;+\;(\om\sqcup \om)$.
	These two \wqos have the same ordinal invariants,
	but $\Pf(A_1)$ and $\Pf(A_2)$ disagree on all 3 ordinal invariants.
\end{example}

\begin{proof}[Proof of Claim]
	Observe that $\Pf(A_1) \cong \om \cdot 2 \times \om \cdot 2$,
	and that $\Pf(A_2) \wpoiso (\om \times \om) + (\om \times \om)$ (\Cref{lem:pf-plus-sqcup}).
	As a consequence, we obtain the following ordinal
        invariants\footnote{In these and later computations, we rely
        on~\cite{abraham87} and \cite{vialard2024} for the width of Cartesian products of ordinals.}:
	\begin{center}
		\begin{tabular}{c c c c}
			\toprule
			\textbf{A} & $\o$            & $\h$          & $\w$          \\ \midrule
			$A_1$      & $\om \cdot 4$   & $\om \cdot 2$ & $2$           \\
			$A_2$      & $\om \cdot 4$   & $\om \cdot 2$ & $2$           \\
			\addlinespace
			$\Pf(A_1)$ & $\om^2 \cdot 4$ & $\om \cdot 3$ & $\om \cdot 3$ \\
			$\Pf(A_2)$ & $\om^2 \cdot 2$ & $\om \cdot 2$ & $\om$         \\
			\bottomrule
		\end{tabular}
	\end{center}
\end{proof}

The non-functionality of the ordinal invariants of $\Pf(A)$ does not prevent
us from building bounds
for $\h$, $\w$, and $\o$.
Our main argument to prove upper and lower bounds will be structural in the
sense that, from the form of the ordinal invariants of $A$ (e.g., are they successor or limit ordinals? finite or infinite? indecomposable?), we will deduce
some structure $B$ related to $A$ such that the invariants of $\Pf(B)$ are easier
to estimate. Here is a first example, a structural lemma used for
the upper bound on the maximal order type:
\begin{restatable}[Sandwich Lemma]{lemma}{motsum}
	\label{lem:mot-sum}
	Let $(A,\le)$ be a \wqo such that $\o(A) = \alpha + \beta$ for some
	ordinals $\alpha, \beta$.
	Then there exists a partition $A = A_\alpha \uplus A_\beta$ of
        $A$ with
	$\o(A_\alpha) = \alpha$, $\o(A_\beta) = \beta$, and such that
	$A_\alpha \dunion A_\beta\leqaug A\leqaug A_\alpha + A_\beta$.
\end{restatable}

\Cref{app:proof:mot-sum} contains the proof of \Cref{lem:mot-sum}, which is a
refinement of a previous result from \cite[Theorem
3.2]{dejongh77}, where the case $\beta=1$ is handled.

\subsection{Maximal Order Type of $\Pf(A)$}
\label{ssec-mot-Pf}

Let $A$ be a \wqo. It is clear that, modulo isomorphism, $1 \lexsum A$
is a
\kl{substructure} of $\Pf(A)$
through the map $0 \mapsto \emptyset$ and, for $x\in A$, $x \mapsto \set{x}$.
Using \Cref{tab:fam:la-famille} and \Cref{lem:invariants-monotonicity},
we obtain the following lower bound:
\begin{equation}
\label{eq-lb-oPf}
1 + \o(A) = \o(1 \lexsum A) \leq \o(\Pf(A))\:.
\end{equation}
For the upper bound,
we will rely on \Cref{lem:mot-sum} to recursively decompose
$A$ and prove that $\o(\Pf(A)) \leq 2^{\o(A)}$.

\begin{therm}
	\label{thm:pf-mot}
	For all \wqo $A$,
	$1+\o(A) \le \o(\Pf(A)) \le 2^{\o(A)}$.
\end{therm}
\begin{proof}
We prove the second inequality by induction on $\o(A)$. We consider three cases:
\begin{itemize}
\item[1.~$\o(A)$ is finite:] Assume w.l.o.g.\ that $A$ is a \wpo. Then
  $\o(A)$ is the cardinal of $A$ and $\o(\Pf(A))$ is less than, or
  equal to, the cardinal of $\Pf(A)$ which is $2^{\o(A)}$.

\item[2.~$\o(A)$ is an infinite successor ordinal:] Then $\o(A) = \alpha
  + n$ with $\alpha$ limit and infinite, and $1\leq n<\om$.  Using
  \Cref{lem:mot-sum} we can split $A$ as $A = A_1 \uplus A_2$ with
  $\o(A_1) = \alpha$, $\o(A_2) = n$, and $A \geqaug A_1 \sqcup A_2$.
    Because $\Pf$ is monotonic with respect to augmentations, we
    see that $\Pf(A) \geqaug \Pf(A_1 \sqcup A_2)$. Then,
    from \Cref{lem:invariants-monotonicity}, we deduce
    that
    $\o(\Pf(A)) \leq \o(\Pf(A_1 \sqcup A_2))$.
    But we know from \Cref{lem:pf-plus-sqcup}
    that $\Pf(A_1 \sqcup A_2) \wpoiso \Pf(A_1) \cartprod \Pf(A_2)$,
    and from \Cref{tab:fam:la-famille} that
    the \kl{maximal order type} of a \kl{Cartesian product} is the
    natural product, denoted $\otimes$,
    of the components' maximal order types,
    hence we obtain:
    \begin{equation*}
	      \o(\Pf(A)) \leq \o(\Pf(A_1) \cartprod \Pf(A_2)) = \o(\Pf(A_1)) \otimes \o(\Pf(A_2))
              \: .
    \end{equation*}
  The induction hypothesis gives $\o(\Pf(A_1)) \leq 2^{\alpha}$ and
  $\o(\Pf(A_2)) \leq 2^n$.
  Therefore, $\o(\Pf(A))\leq 2^\alpha\otimes 2^n= 2^{\alpha + n}$.
  We have proven that
  $\o(\Pf(A)) \leq 2^{\o(A)}$ in this case.

\item[3.~$\o(A)$ is a limit ordinal:] We use the descent equations:
   \begin{equation}
  \label{eq-proof-pf-mot-reso}
         \o(\Pf(A)) \eqby{\eqref{eq-Des-o}} \sup_{S \in \Pf(A)} \bigl(\o(\Pf(A)_{\not\hoaregeq S}) + 1\bigr)\:.
   \end{equation}
   Given a finite set $S \in \Pf(A)$,  we further decompose the set $\Pf(A)_{\not\hoaregeq S}$:
   \begin{align*}
       \Pf(A)_{\not\hoaregeq S} & = \{ T \in \Pf(A) \mid S \not\hoareleq T \}
     \\
     & = \{ T \in \Pf(A) \mid \exists x \in S, \forall y \in T, x \not\le y \}
     \\
     & = \{ T \in \Pf(A) \mid \exists x \in S, T \in \Pf(A_{\not\geq x}) \}
     \\
    & = \bigcup_{x \in S} \Pf(A_{\not\geq x})\:.
   \end{align*}
    As a consequence, $\Pf(A)_{\not\hoaregeq S}$ is an \kl{augmentation} of
    the \kl{disjoint sum} $\bigsqcup_{x \in S} \Pf(A_{\not\geq x})$, and
    the following inequality holds with reasoning similar to
    the successor case (using
    \Cref{lem:invariants-monotonicity,lem:pf-plus-sqcup,tab:fam:la-famille}):
   \begin{equation}
   \label{eqn:mot-pf-bigoplus}
        \o(\Pf(A)_{\not\hoaregeq S}) \leq \bigoplus_{x \in S} \o(\Pf(A_{\not\geq x}))
                        \leq \bigl[\max_{x\in S} \o(\Pf(A_{\not\geq x})) \bigr]\otimes|S|
\;.
   \end{equation}
   Let us write $\beta = \max_{x \in S} \o(A_{\not\geq
   x})$. Eq.~\eqref{eq-Des-o} entails $\beta < \o(A)$.
   By induction hypothesis, for all $x\in S$, we have
   $\o(\Pf(A_{\not\geq x})) \leq 2^{\o(A_{\not\geq x})} \leq
   2^{\beta}$. Hence
   \begin{align*}
        \o(\Pf(A)_{\not\hoaregeq S}) &\leq 2^{\beta} \otimes |S| < 2^{\o(A)}\:,
   \end{align*}
   where the last step relies on $2^{\o(A)}$ being \kl{additively indecomposable}
   when $\alpha$ is limit.
   Using again this fact that $2^{\o(A)}$ is \kl{additively indecomposable},
   we see that $\o(\Pf(A)_{\not\hoaregeq S}) + 1 < 2^{\o(A)}$. Since
   this holds for any $S$, Eq.~\eqref{eq-proof-pf-mot-reso} entails
   $\o(\Pf(A)) \le \o(A)$. \qedhere
\end{itemize}
\end{proof}

Notice that in the proof of \Cref{thm:pf-mot}, we relate the
residual of a set $S$ in $\Pf(A)$ with the residuals $A_{\not\geq x}$
of the elements of $S$ in $A$. This method will be used several times
in this article.

\subsection{Height of $\Pf(A)$}

Let $A$ be a \wqo. For the same reason as for
Eq.~\eqref{eq-lb-oPf}, that is, because $1 \lexsum A$ is a
\kl{substructure} of $\Pf(A)$, we have the following lower bound:
\begin{equation}
\label{eq-lb-hPf}
                    1 + \h(A) \leq \h(\Pf(A)) \:.
\end{equation}

Before going further, we would like to produce a non-tight upper bound for
$\h(\Pf(A))$, that relates it to the height of $\Pfull(A)$ (recall that the
height is defined whenever the order is well-founded). While crude, this bound
will serve in \Cref{sec:algebra-well-behaved}.

\begin{remark}
\label{rem-Pinf}
  $\Pf(A)$ is a \kl{substructure} of the full powerset $\Pfull(A)$ ordered with
  \kl{Hoare's embedding}, which is \kl{{\wpo}-isomorphic} to
  $(\Down(A),\subseteq)$, the  downward-closed
  subsets of $A$ ordered by inclusion. Therefore, $\h(\Pf(A))\leq
  \h(\Down(A))=\o(A)+1$ \citep[Theorem 3.5]{dzamonja2020}.
\end{remark}
One can improve this upper bound.
As in the proof of \Cref{thm:pf-mot}, we rely on a
structural decomposition lemma to handle
the successor case,
and on the relationship between the residuals
of $\Pf(A)$ and those of $A$ for the limit case.
Since the structural lemma is simpler to obtain, we include its proof.

\begin{restatable}{lemma}{hsum}
	\label{lem-pf-height-destruct}
  Let $A$ be a \wqo such that $\h(A)=\alpha+1$. Then there exists a partition
  $A=A_\bot \uplus A_\top\leqaug  A_\bot \lexsum A_\top$ such that
  $\h(A_\bot)=\alpha$ and $A_\top$ is \kl{{\wpo}-isomorphic} to an \kl{antichain}.
\end{restatable}
\begin{proof}
Without any loss of generality, let us assume that $A$ is a \wpo.
In this proof we use the same result of \cite{wolk67}
that was used in \Cref{lem:desc-eq}.
The \kl{rank function} $\orank$
of $\DecSeq(A)$ can be used to associate a rank with every element of $A$
as follows:
  the rank
of $x$ is $\orank((x))$, referring to $(x)$, the decreasing sequence of
length $1$ built with $x$.

Observe that elements of equal rank always form an
  antichain, and that the direct sum of the antichains obtained from the partition
  of $A$ by rank is an augmentation of $A$.

  If we now let $A_\top = \setof{x\in A}{\orank((x))=\alpha}$ and
	$A_\bot = A \setminus A_\top$,
  then $\h(A_\bot)=\alpha$ and $A_\top$ is an antichain, which
  concludes the proof.
\end{proof}

\begin{therm}
	\label{thm:pf-height}
	For all \wqo $A$, there exists $m \in \Nat$ such that,
	\begin{equation*}
		1 + \h(A) \leq \h(\Pf(A)) \leq
		\begin{cases}
			2^{\h(A)}     & \text{ when } \h(A) \text{ is a limit ordinal,}     \\
			2^{\h(A)}\cdot m & \text{ when } \h(A) \text{ is a successor ordinal.} \\
		\end{cases}
	\end{equation*}
\end{therm}
\begin{proof}
There only remains to prove the upper bound. We do this by induction on $\h(A)$.
	\begin{itemize}
		\item[If $\h(A)$ is finite,] then
		      because $A$ is a \wqo, it is {\wpo}-isomorphic
                      to a  finite set, and so
		      $\Pf(A)$ is too.
		      As a consequence, $\h(\Pf(A)) \leq m$
		      for some $m \in \Nat$.
		\item[If $\h(A) = \alpha +1$] is an infinite successor
                  ordinal, we rely on
		      \Cref{lem-pf-height-destruct}
		      and pick a  decomposition
		      $A = A_\bot \uplus A_\top$ such that
		      $\h(A_\bot) = \alpha$
          and $A_\top \wpoiso \FAnt{m}$ for some $1\leq m<\om$.
		      Without loss of generality,
              assume that $A$ is a \wpo.
              Recall that $\Pf(A)\wpoiso\Ant(A)$, the \wpo of
              antichains of $A$ ordered by Hoare's embedding and observe that any antichain $S$ of $A$ can be decomposed as an antichain of $S_\bot$ of $A_\bot$ and an antichain $S_\top$ of $A_\top$. Let $S,S'$ be two antichains of $A$. Then $S\hoareleq S'$ only if either $S_\top \hoarele S'_\top$,
              or $S_\top = S'_\top$ and $S_\bot \hoareleq S'_\bot$. 
          Hence,
          antichains of $A$ can be seen
          as \kl{lexicographically ordered pairs} of
          antichains of $A_\top$ and antichains of $A_\bot$:

		      \begin{align*}
		      \Ant(A)&\leqaug \setof{(S_\bot,S_\top)\in\Ant(A_\bot)\cdot\Ant(A_\top)}{S_\bot\cup S_\top\in \Ant(A)}\\
		      &\leqstruct \Ant(A_\bot)\lexprod \Ant(A_\top)\;.
		      \end{align*}
          Because $\h$ is monotonic with respect to
          \kl{augmentations} and \kl{substructures}
          for \wpos (\Cref{lem:invariants-monotonicity}),
          we conclude that
          $\h(\Ant(A))\leq \h(\Ant(A_\bot)\lexprod \Ant(A_\top))$, with
          \Cref{tab:fam:la-famille} giving
          $\h(\Ant(A_\bot) \lexprod \Ant(A_\top)) =
           \h(\Ant(A_\bot))\cdot \h(\Ant(A_\top))$.
          From the induction hypothesis,
          there exists $m \in \Nat$ such that
          $\h(\Ant(A_\bot)) \leq 2^{\alpha}\cdot m$ for some $m$,
          and $\h(\Ant(A_\top)) = \h(\FAnt{m'})$ for some $m'$.
          Therefore, again using \Cref{tab:fam:la-famille},
          $\h(\Ant(A)) \leq 2^{\alpha}\cdot m \cdot \h(\FAnt{m'})$.
          We conclude, as desired, that
          $\h(\Pf(A)) \leq 2^{\alpha} \cdot (m \times (1 + m'))$.

		\item[If $\h(A) = \alpha$ where $\alpha$ is limit ordinal,]
		      then using Eq.~\eqref{eq-Des-h}:
		      \begin{equation*}
			      \h(\Pf(A)) = \sup_{S \in \Pf(A)}
                              \bigl(\h(\Pf(A)_{\hoarele S}) + 1\bigr) \:.
		      \end{equation*}
		      Let us fix $S \in \Pf(A)$ and
		      let $B \defined \Pf(A)_{\hoarele S}$. Then
		      \begin{align*}
			      B & = \{ T \in \Pf(A) \mid T \hoarele S \}                    \\
			        & \subseteq \{ T \in \Pf(A) \mid T \subseteq \dwc S \} \\
 			        & = \Pf(\dwc S)     \:.
		      \end{align*}
We recall that $\dwc S\eqdef \bigcup_{x\in S} A_{\leq x}$.
		      Because $\h(A) = \alpha$ is a limit ordinal, for all $x\in A$, $\h(A_{\leq x}) = \h(A_{<x})+1<\alpha$. Since $S\subseteq A$ is finite,
		      $\h(\dwc S) < \alpha$.

		      We can therefore apply the induction hypothesis:
		      There exists $m \in \Nat$
		      such that
		      $\h(B) \leq 2^{\h(\dwc S)} < 2^{\alpha}$ if
                      $\h(\dwc S)$ is limit,
		      and $\h(B) \leq 2^{\h(\dwc S)}\cdot m < 2^{\alpha}$ otherwise.

		      This proves that $\h(\Pf(A))\leq 2^{\alpha}$.
		      \qedhere
	\end{itemize}
\end{proof}

Note that the upper bound of \Cref{thm:pf-height} could be refined further to
control the constant $m$ in the successor case. To that end, one could change
the induction so that in the successor case, one decomposes $\h(A)$ as $\alpha
+ n$ with $\alpha$ limit, and then see that $m$ is bounded by
$\h(\Pf(\FAnt{n}))$. We refrained from doing so since this would make
both the proof and the statement much more complex, with little
added value,

\subsection{Width of $\Pf(A)$}

Thanks to \Cref{thm:pf-mot}, and since $\w\leq\o$, we  see that $\w(\Pf(A))\leq
2^{\o(A)}$. This inequality does not bound the width of $\Pf(A)$ by an
expression depending on $\w(A)$, and we now claim that there exists no such
upper bound, anticipating the tightness results of \Cref{sec:tightness}.

\begin{example}
\label{exmp:w-low-2}
	For all infinite ordinals $\alpha$, let $A_\alpha=\om \sqcup \alpha$.
	Then $\w(A_\alpha) = 2$ and $\w(\Pf(A_\alpha)) = \alpha$.
\end{example}
\begin{proof}
\Cref{lem:pf-plus-sqcup} gives $\Pf(A_\alpha)\cong
\Pf(\om)\times\Pf(\alpha)\cong
\om\times \alpha$.  There remains to compute the width of a Cartesian
product of ordinals:
     $\w(\om\times\alpha)=\alpha$ \citep{abraham87,vialard2024}.
\end{proof}

As a consequence, we will focus on providing a lower bound for the width
of $\Pf(A)$.
As opposed to the case of $\h$ and $\o$, this
lower bound happens to be non-trivial to obtain.
Our first
goal is to provide structural lemmas analogous to
\Cref{lem:mot-sum,lem-pf-height-destruct}.

\begin{restatable}{lemma}{structuralwidth}
	\label{lem:successor-width}
	Let $A$ be a \wqo such that $\w(A) = \alpha + n$
	with $0\leq n < \om$.
	Then there exists $B$ such that
	$B \sqcup \FAnt{n} \leqstruct A$ and  $\w(B)=\alpha$.
\end{restatable}
\begin{proof}
By induction on $n$. We build an antichain $X=\{x_1,\ldots,x_n\}$ of
$A$ such that $\w(A_{\perp X})=\alpha$ so letting $B=A_{\perp X}$
proves the claim.  The case $n=0$ is trivial.

When $n>0$, and according to Eq.~\eqref{eq-Des-w}, $\w(A) = \sup_{x \in
A} (\w(A_{\perp x}) + 1)$, therefore there exists $x\in A$ such that
$\w(A_{\perp x}) = \alpha + (n-1)$. By ind.\ hyp., there is an
antichain $X'=\{x_1,\ldots,x_{n-1}\}$ of $A_{\perp x}$ such that
$\w(A_{\perp x\cup X'})=\alpha$. Now, and since $X'$ is in $A_{\perp
x}$, letting $X=\{x\}\cup X'$ yields an antichain and concludes the
proof.
\end{proof}

\begin{corollary}\label{cor:gamma-k-substruct}
	Let $A$ be a \wqo. If $\w(A)=k < \om$, then
	$\FAnt{k} \leqstruct A$.
\end{corollary}

A lower bound using the structural decomposition lemmas above will naturally
rely on computing widths of the form $\w(\Pf(B \sqcup \FAnt{k}))$, which can be
rewritten as $\w(\Pf(B) \times \Pf(\FAnt{k}))$ according to
\Cref{lem:pf-plus-sqcup}. One major roadblock to such a proof scheme is that
\Cref{tab:fam:la-famille} does not provide a closed expression for the width of
a Cartesian product. To tackle this issue, we leverage the work of \cite{dzamonja2020}
and \cite{vialard2024}.

\begin{restatable}[Lower bound on the width of the Cartesian product]{lemma}{lembiftransferable}
	\label{lem-isa}
  Let $A,B$ be two \wqos such that $\w(B)\geq\om$ is \kl{additively indecomposable}.
	Then $\w(A \cartprod B)\geq \w(B) \cdot \o(A)$.
\end{restatable}
Since the proof of \Cref{lem-isa} uses technical notions that are not central
to this article, we refer the reader to \Cref{app:proof:lem-isa} for details.

In the following bounds for the width of the finitary powerset, we abstract
away the finite constants (as in \Cref{thm:pf-height}) to simplify the analysis.
Note that the finite parts are computed as
$\w(\Pf(\FAnt{k}))$ for some $k$, which is $\binom{k}{\lfloor k/2\rfloor}$
by Sperner’s theorem.
\begin{therm}
	\label{thm:pf-width}
	Let $A$ be a \wqo. If $\w(A)$ is infinite then $\w(\Pf(A)) \ge
        2^{\w(A)}$.
        If $\w(A)=k<\om$ then $\w(\Pf(A)) \ge \w(\Pf(\FAnt{k}))$.
\end{therm}
\begin{proof}
	We prove the result by induction on $\w(A)$.
	\begin{itemize}
		\item[If $ \w(A)=k < \om$] then
          by \Cref{cor:gamma-k-substruct},
		       $\FAnt{k}\leqstruct A$,
           hence by monotonicity of $\Pf$,
                       $\Pf(\FAnt{k})\leqstruct \Pf(A)$.
          Since the \kl{ordinal width} is monotone with respect
          to \kl{substructures} (\Cref{lem:invariants-monotonicity}),
          we conclude that
		       \begin{equation*}
                   \w(\Pf(\FAnt{k})) \leq \w(\Pf(A)) \: .
               \end{equation*}
		\item[If $\w(A) = \alpha + n$ with $\alpha$ limit and
                $1\leq n<\om$]
		      then by
		      \Cref{lem:successor-width}, there exists a \wqo $B$ of width $\alpha$,
		      such that $B \sqcup \FAnt{n} \leqstruct A$.
		      Hence $\w(\Pf(A))\geq \w(\Pf(B \sqcup
                      \FAnt{n}))=\w(\Pf(B) \times \Pf(\FAnt{n})) $
                      according to
                      \Cref{lem:invariants-monotonicity,lem:pf-plus-sqcup}.
		     By induction hypothesis, $2^\alpha \leq \w(\Pf(B))$.
		      Therefore, by \Cref{lem:desc-eq}, there exists
		      $C \leqstruct \Pf(B)$ such that $\w(C) = 2^\alpha$.
          The latter is an \kl{additively indecomposable} ordinal
		      since $\alpha$ is limit and infinite, hence
                      we may continue with
		      \begin{align*}
			      \w(\Pf(A)) &\geq \w(\Pf(B) \times \Pf(\FAnt{n}))
			      \geq \w(C \times \Pf(\FAnt{n})),
      \\
                      \shortintertext{and with \Cref{lem-isa}}
			      &\geq \w(C)\cdot \o(\Pf(\FAnt{n})) =  2^\alpha \cdot 2^n = 2^{\alpha + n},
		      \end{align*}
		      which is the expected lower bound.
		\item[If $\w(A) = \om$]
				then for all $k<\om$, $\w(\Pf(A))\geq
                                \w(\Pf(\FAnt{k}))$. Observe that
				$\sup_{k<\om} \w(\Pf(\FAnt{k}))=\om$,
                                and therefore $\w(\Pf(A))\geq
                                \om=2^\om$.
		\item[If $\w(A) = \alpha>\om$ and $\alpha$ is a limit ordinal,]
		      then for all $\om<\beta < \alpha$, there exists
                      (\Cref{lem:desc-eq})
		      a $B_\beta \leqstruct A$ such that $\w(B_\beta) = \beta$.
		      By induction hypothesis, and since $\om<\beta$,
		      $2^\beta \leq \w(\Pf(B)) \leq \w(\Pf(A))$.
		      Therefore,
              \begin{equation*}
                  \w(\Pf(A)) \geq \sup_{\beta < \alpha} 2^\beta = 2^\alpha
                  \: .
                  \qedhere
              \end{equation*}
	\end{itemize}
\end{proof}

\section{Tightness of the Bounds}
\label{sec:tightness}

The upper and lower bounds provided in \Cref{sec:up-lower-bounds} allow huge
gaps. For instance, when $\h(A) = \om^2$, \Cref{thm:pf-height} shows that
$\h(\Pf(A))$ can lie between $1+\om^2$, i.e., $\om^2$, and $2^{\om^2}$, i.e.,
$\om^\om$. We proceed to prove that these gaps cannot be avoided by exhibiting,
for every measure $\someInvariant \in \set{\w, \o, \h}$, for upper and lower
bounds, a family  $(X_\alpha)_\alpha$  of \wqos such that
$\someInvariant(X_\alpha) = \alpha$ and such that
$\someInvariant(\Pf(X_\alpha))$ equals the provided (upper or lower) bound.

\subsection{Tight lower bound for $\w(\Pf(A))$ and upper bound for $\o(\Pf(A))$}
\label{sec-halpha}

We first introduce a family of \wqos $(\FPhi{\alpha})_\alpha$ that will reach
both the lower bounds for $\w(\Pf(X))$ and the upper bound for $\o(\Pf(X))$. We
recall that,  for any $k<\om$, $\FAnt{k}$ denotes a \wqo of $k$ incomparable
elements. The definition of $\FPhi{\alpha}$ will use the \kl{Cantor
normal form} (CNF) of $\alpha$, i.e., the unique way of writing $\alpha$ as
$\alpha = \om^{\alpha_1} + \om^{\alpha_2} + \cdots + \om^{\alpha_n}$ with
$\alpha_1 \ge \alpha_2 \ge \cdots \ge \alpha_n$.

\begin{definition}
    \label{def:fphi}
We first define inductively the $\FPhi{\alpha}$ for $\alpha$ a power
of $\om$:\\
-- $\FPhi{\om^0} \eqdef \FAnt{1}$,
\\
-- $\FPhi{\om^1} \eqdef \sum_{n\in \om} \FAnt{n} = \FAnt{0}+\FAnt{1}+\FAnt{2}+\cdots$, see
\Cref{fig:H-om-1},
\\
-- for $\alpha > 0$, $\FPhi{\om^{\omega^\alpha}} = \sum_{\beta <
\omega^\alpha} \FPhi{\om^{\beta}}$,
\\
-- and for $\alpha=\om^{\alpha_1} + \dots +
                  \om^{\alpha_n}$ in \kl{Cantor normal form}, with $n>1$,
		      $\FPhi{\om^{\alpha} }
			      \eqdef \FPhi{\om^{\om^{\alpha_1}}}
              \lexprod \FPhi{\om^{\om^{\alpha_2}}}
              \cdots \lexprod
                  \FPhi{\om^{\om^{\alpha_n}}}$
              (the \kl{lexicographic product} of the family of
              $\FPhi{\om^{\om^{\alpha_i}}}$).
\end{definition}

\begin{figure}[ht]
\begin{center}
\begin{tikzpicture}[auto,x=5mm,y=3mm,anchor=mid,baseline,node distance=1.3em]
{%
\def\Xmid{0}%
\def\Gcol{6}%
\def\Xd{1}%
\def\Ybot{-5}%
\def\Ylow{0}%
\def\Yup{4}%
\def\Ymidb{{0.5*(\Ylow+\Ybot)}}%
\def\Ymid{{0.5*(\Yup+\Ylow)}}%
\def\Yd{1.8}%

\foreach \i in {0,...,3} \node at ({(\i-1.5)*\Xd+\Xmid},\Ylow) {$\bullet$};
\node at (\Gcol,\Ylow) {$\FAnt{4}$};
\foreach \i in {0,...,4} \node at ({(\i-2.0)*\Xd+\Xmid},\Yup) {$\bullet$};
\node at (\Gcol,\Yup) {$\FAnt{5}$};
\foreach \i in {0} \node at ({\i*\Xd+\Xmid},\Ybot) {$\bullet$};
\node at (\Gcol,\Ybot) {$\FAnt{1}$};
\node[rotate=90] at (\Xmid,\Ymid) {$\bm{<}$};
\draw[thick,decorate,decoration={brace,raise=-2.2ex}] ({\Xmid-1.7*\Xd},\Ymid) -- ({\Xmid+1.7*\Xd},\Ymid);
\draw[thick,decorate,decoration={brace,mirror,raise=-2.2ex}] ({\Xmid-2.2*\Xd},\Ymid) -- ({\Xmid+2.2*\Xd},\Ymid);
\node at (\Xmid,\Ymidb) {$\bm{\vdots}$};
\node at (\Xmid,\Yup+\Yd) {$\bm{\vdots}$};
\node at (\Gcol,\Ymidb) {$\bm{\vdots}$};
\node at (\Gcol,\Yup+\Yd) {$\bm{\vdots}$};

}%
\end{tikzpicture}
\end{center}
\caption{$\FPhi{\om^1}\eqdef \FAnt{0} + \FAnt{1} + \FAnt{2} + \cdots$}
\label{fig:H-om-1}
\end{figure}

\begin{restatable}{proposition}{canofam}
	\label{prop:cano-fam}
	For all ordinals $\alpha$:
	\[
		\h(\FPhi{\om^{\alpha}}) = \w(\FPhi{\om^{\alpha}}) = \o(\FPhi{\om^{\alpha}}) = \om^\alpha.
	\]
\end{restatable}

The proof of \Cref{prop:cano-fam} is done by induction on $\alpha$ and follows
from \Cref{tab:fam:la-famille}, except that some technical care is needed to
handle the \kl{lexicographic product} case. Therefore, the full proof can be
found in \Cref{app:proof:prop:cano-fam}.

Unfortunately, the family $\FPhi{\om^\alpha}$ only covers ordinals of the form
$\om^\alpha$. We now extend it to all ordinals as follows: for an ordinal
$\alpha$ in \kl{Cantor normal form} $\om^{\alpha_1} + \cdots + \om^{\alpha_n}$,
we let $\FPhi{\alpha}$ be defined as $\FPhi{\om^{\alpha_1}} \dunion \cdots
\dunion \FPhi{\om^{\alpha_n}}$. While this ensures $\w(\FPhi{\alpha}) =
\o(\FPhi{\alpha}) = \alpha$ for all $\alpha$, observe that $\FPhi{\alpha}$ has
height $\max_i \om^{\alpha_i}$ and not $\alpha$.

There remains to evaluate the
ordinal invariants of $\Pf(\FPhi{\alpha})$.
\begin{therm}
	\label{thm:pf-o-ub-tight}
	\label{thm:pf-w-lb-tight}
	For every $\alpha$,
	$\o(\FPhi{\alpha}) = \w(\FPhi{\alpha})=\alpha$. Furthermore,
	$\w(\Pf(\FPhi{\alpha}))$ attains the lower bound of \Cref{thm:pf-width} and $\o(\Pf(\FPhi{\alpha}))$ attains the upper bound
	of \Cref{thm:pf-mot}.
\end{therm}
\begin{proof}
	From \Cref{thm:pf-width,thm:pf-mot} we conclude that for $\alpha$ infinite:
	\begin{equation*}
		2^\alpha = 2^{\w(\FPhi{\alpha})}
		\le \w(\Pf(\FPhi{\alpha}))
		\le \o(\Pf(\FPhi{\alpha}))
		\le 2^{\o(\FPhi{\alpha})} = 2^\alpha\:.
	\end{equation*}
	It follows that $\w(\Pf(\FPhi{\alpha})) =
        \o(\Pf(\FPhi{\alpha})) = 2^\alpha$. Hence, the upper bound in
        \Cref{thm:pf-mot} and the lower bound in
        \Cref{thm:pf-width} are attained.

	Lastly, for $\alpha=k$ finite, $\FPhi{k}=\FAnt{1}\dunion
        \cdots\dunion\FAnt{1}\cong\FAnt{k}$. Now $\w(\Pf(\FAnt{k}))$ is precisely the lower bound given in
        \Cref{thm:pf-width}, while
        $(\Pf(\FAnt{k}),\hoareleq) = (\Pf(\FAnt{k}),\subseteq)$
        is a \wpo of cardinal $2^k$, hence of
        \kl{maximal order type} $2^k$ too.
		      \qedhere
\end{proof}

\subsection{Tight upper bounds for $\h(\Pf(A))$}
\label{sec-palpha}

We now provide a family of \wqos reaching the upper bound for the
height. For this recall that \kl{indecomposable} ordinals are of the form $\om^{\om^\beta}$.

\begin{definition}
	\label{def:f-sim-alpha}
	For $\alpha=\om^{\om^\beta}$ indecomposable and $n,k<\omega$, define
        $\FSim{\alpha}$ and
$\FSim{\alpha+n,k}$ as follows:\\
\begin{align*}
\FSim{\alpha} &\defined \begin{cases}\om &\text{ if $\alpha=\omega$,
i.e., $\beta=0$,}\\
                         \Pf(\alpha^\seq) &\text {otherwise,}
                        \end{cases}
\\
  \FSim{\alpha+n,k}&\defined \bigl[(\FSim{\alpha}) + n\bigr]\times
  \FAnt{k}
                       \cong  \bigsqcup_{i=1}^k (\FSim{\alpha}) + n.
\end{align*}
\end{definition}
For $\alpha$ indecomposable, the upper bound given in
\Cref{thm:pf-height} is matched.
\begin{lemma} 
\label{lem:h-FSim}
$\h(\FSim{\alpha})=\alpha$ and
$\h(\Pf(\FSim{\alpha}))=2^\alpha$.
\end{lemma}
\begin{proof}
Indeed, for $\alpha=\omega$, the claim holds thanks to $2^\om=\om$ and
$\Pf(\om )\wpoiso \om$. For $\alpha>\omega$, $\FSim{\alpha}$ is an
\kl{elementary \wqo} (as defined in the next section).  We will see
in \Cref{sec:elementary-height} that $\h(\FSim{\alpha}) = \alpha$, and
$\h(\Pf(\FSim{\alpha})) = 2^\alpha$. 
\end{proof}
Now, for some successor $\alpha+n$ with $\alpha$ indecomposable,
the upper bounds given by \Cref{thm:pf-height}
can be matched for any $n>0$.
\begin{lemma} 
\label{lem:h-FSim-m}
For $\alpha$ indecomposable and $0<n,m<\omega$, let
$k=2^n\cdot m$. Then
    \begin{equation*}
    \h(\FSim{\alpha+n,k}) = \alpha + n
    \text{ and }
    \h(\Pf(\FSim{\alpha+n,k})) \geq 2^{\alpha+n} \cdot m
    \: .
    \end{equation*}
\end{lemma}
\begin{proof}
Since, modulo isomorphism, $\FSim{\alpha+n,k}$ is $\bigsqcup_{i=1}^k
(\FSim{\alpha}) + n$,
we obtain $\h(\FSim{\alpha+n,k}) = \h((\FSim{\alpha}) + n) = \alpha +n$
by relying on \Cref{tab:fam:la-famille} and \Cref{lem:h-FSim}.
With \Cref{lem:pf-plus-sqcup}, we also obtain
$\Pf(\FSim{\alpha+n,k})
\cong (\Pf((\FSim{\alpha}) + n))^{\times k}
\wpoiso(\Pf(\FSim{\alpha}) + n)^{\times k}$, and 
hence $\h(\Pf(\FSim{\alpha+n,k}))=\h[\Pf(\FSim{\alpha}) + n] \hplus
\cdots \hplus \h[\Pf(\FSim{\alpha}) + n]= (2^{\alpha}+n) \hplus \dots
\hplus (2^{\alpha}+n)\geq 2^\alpha\cdot k = 2^{\alpha+n}\cdot m$.
\end{proof}

\section{An Algebra of Well-Behaved Wqos}%
\label{sec:algebra-well-behaved}

\AP The results in \Cref{sec:up-lower-bounds,sec:tightness} show that one
cannot derive the height, width or maximal order type of $\Pf(A)$ from the
invariants of $A$. Hence the powerset construction behaves differently from
other operations for which ordinal invariants can be computed in a
compositional way (see \Cref{tab:fam:la-famille}).

In this section we show that the situation is not so negative. We consider a
rather large family of ``elementary'' \wqos closed under most classical
operations, for which we can compute the ordinal invariants of $\Pf(A)$, even
in the presence of nested powerset constructions. This extends previous results
in the literature for families of \wqos which do not include the \kl{finitary
powerset} construction~\citep{vialard2024}. Most notably, the family of
\kl{elementary \wqos} \emph{excludes} the \kl{lexicographic sum}, that behaves
badly with respect to the \kl{Cartesian product} as noticed by~\cite{vialard2024}.

\begin{definition}[Elementary \wqo]
	\label{def:fam:elementary-wqo}
  The family of \intro{elementary \wqos}
	is given by the following abstract grammar
	\begin{equation*}
		E \defined \alpha\geq\omom \text{ \kl{indecomposable}}
		\mid E_1 \dunion E_2
		\mid E_1 \times E_2
		\mid E^\seq
		\mid \Mul(E)
		\mid \Pf(E)  .
	\end{equation*}
(These expressions are interpreted as \wqos according to the definitions given in \Cref{sec:basics}.)
\end{definition}

In most cases, the invariants of elementary \wqos can be computed using
\Cref{tab:fam:la-famille}.
However some specific cases, for instance computing the ordinal invariants of $\Pf(E)$, or
computing $\w(E_1 \times E_2)$, have to be handled differently.
To reduce the boilerplate code to a minimum, we  first
normalize our expressions so that we can minimize the
appearance of problematic subexpressions with the help of some isomorphisms
in the spirit of those
described in \Cref{lem:pf-plus-sqcup}. The rewriting
rules given in \Cref{fig:rewrite-rules} preserve \wqos modulo \kl{\wpo-isomorphism},
and define a strongly normalizing, confluent rewrite system on the
expressions for elementary \wqos.

\begin{figure}[ht]
\begin{align*}
	\Pf(\alpha)         & \rightarrow \alpha\;,                        \\
	E\times (E_1\sqcup E_2) & \rightarrow (E\times E_1)\sqcup (E\times E_2)\;, \\
	(E_1\sqcup E_2)\times E & \rightarrow (E_1\times E)\sqcup (E_2\times E)\;, \\
	\Mul(E_1\sqcup E_2)     & \rightarrow \Mul(E_1)\times \Mul(E_2)\;,         \\
	\Pf(E_1\sqcup E_2)      & \rightarrow \Pf(E_1)\times \Pf(E_2)\;.
\end{align*}
\caption{Rewrite rules for elementary wqos}%
\label{fig:rewrite-rules}
\end{figure}

The normal form of an expression is computable. We therefore assume that an
elementary \wqo is always given via its expression  in \emph{normal form},
i.e., that cannot be rewritten by the above rules. Note that the rewrite rule
$\Pf(\alpha) \rightarrow \alpha$ is valid because
\Cref{def:fam:elementary-wqo} requires $\alpha$ to be \kl{indecomposable},
in which case \Cref{lem:pf-plus-sqcup} leads to $\Pf(\alpha) \wpoiso 1 +
\alpha = \alpha$.

\subsection{Maximal order type and width of elementary wqos}%

In this section, we  provide an algorithm to compute
both the width and the maximal order type
of an elementary \wqo $E$.
The main idea of this double computation is that
most elementary \wqos verify the property $\w(E) = \o(E)$,
in which case the following ``sandwich lemma'' can be applied.

\begin{lemma}[Powerset Sandwich]
	\label{pf-sandwich}
	Let $A$ be a \wqo such that $\w(A) = \o(A)\geq \om$. Then
	$\w(\Pf(A))=\o(\Pf(A))=2^{\o(A)}$.
\end{lemma}
\begin{proof}
  According to \Cref{thm:pf-width} for the lower bound, and \Cref{thm:pf-mot}
  for the upper bound, we have
	\begin{equation*}
		2^{\w(A)}\leq
		\w(\Pf(A))\leq\o(\Pf(A))\leq 2^{\o(A)}\;. \qedhere
	\end{equation*}
\end{proof}

Now, some elementary \wqos do not verify $\o(E)=\w(E)$ (take for instance
$E=\alpha$). Moreover, if $E$ is a \kl{Cartesian product}, we cannot easily
check whether $\o(E)=\w(E)$ as \kl(ordinal){widths} of \kl{Cartesian products}
are not yet completely understood. We tackle this issue with recent work from
\cite{vialard2024}, which reduces the \kl(ordinal){width} of a \kl{Cartesian
product} to its \kl{maximal order type} in specific situations.

\begin{therm}[{\citet[Theorem 5.2]{vialard2024}}]
	\label{thm:fam:isa-product}
	\label{thm-app}
	Let $A_1,\dots,A_n$ be \wqos.
	If there exist $i\neq j \in [1,n]$ and $\alpha,\beta>0$ such that $\o(A_i)=\om^{\om\cdot\alpha}$ and $\o(A_{j})=\om^{\om\cdot\beta}$,
	then $\w(A_1\cartprod \cdots\cartprod A_n) = \o(A_1\cartprod\cdots\cartprod A_n)$.
\end{therm}

As for the powerset, it will become clear in \Cref{lem:elem-disjunction} that
most products of \kl{elementary \wqos} satisfy the hypotheses of
\Cref{thm:fam:isa-product}, and that the only exceptions for \kl{elementary
\wqos} given in normal form are either ordinals, on which the \kl{finitary
powerset} construction acts trivially, or  \kl{disjoint sums} of
\kl{elementary \wqos}, on which the \kl{finitary powerset} acts as a
product of \kl{finitary powersets} (see \Cref{lem:pf-plus-sqcup}).

\begin{lemma}[Case disjunction]
	\label{lem:elem-disjunction}
	Let $E$ be an elementary \wqo (given through its expression in
        normal form). Then either:
	\begin{compactenum}[(i)]
	  \item  $\w(E)=\o(E)=\om^{\om\cdot\beta}$ for some ordinal $\beta>0$, or
          \item  $E= E_1 \sqcup E_2$ with $E_1,E_2$ \kl{elementary \wqos}, or
          \item  $E=\alpha\geq\omom$ \kl{indecomposable}.
	\end{compactenum}
\end{lemma}
\begin{proof}
	By induction on the expression in normal form of $E$:
	\begin{itemize}
		\item[Case $E=\alpha$:]
		     then $E$ satisfies $(iii)$.

		\item[Case $E=E_1\sqcup E_2$:] then $E$ satisfies $(ii)$.

		\item[Case $E=E_1^\seq$:] then $E$ satisfies $(i)$
		      (see \Cref{tab:fam:la-famille} and note that
                  $\o(E_1)>1$ for elementary $E_1$).

		\item[Case $E=\Mul(E_1)$:] By induction hypothesis $E_1$ satisfies either $(i)$ or $(iii)$
      (not $(ii)$ since $E$ is in normal form).
      Hence $\o(E_1)>1$ is \kl{additively indecomposable}, therefore $E$ satisfies $(i)$, see \Cref{tab:fam:la-famille}.

		\item[Case $E=E_1\times E_2$:]
		Here too $E_1$ and $E_2$ satisfy either $(i)$ or $(iii)$.
		      Thus $\o(E_1)$ and $\o(E_2)$ are  $\om^{\om\cdot\beta_1}$ and $\om^{\om\cdot\beta_2}$ for some $\beta_1,\beta_2>0$. Therefore, according to \Cref{thm-app}, $\w(E)=\o(E)=\o(E_1)\otimes \o(E_2)=\om^{\om\cdot(\beta_1\oplus\beta_2)}$. Finally $E$ verifies $(i)$.

		\item[Case $E=\Pf(E_1)$:]
		the induction hypothesis and the fact $E$ is in normal
                form entail that $E_1$ satisfies $(i)$, and then
                $2^{\o(E_1)}=\om^{\o(E_1)}$ (recall \Cref{fact:two-exponentiation}).
                Thus, with \Cref{pf-sandwich},
                $\w(E)=\o(E)=2^{\o(E_1)}=\om^{\o(E_1)}$ and hence $E$ satisfies $(i)$.
	\end{itemize}
\end{proof}
The proof of \Cref{lem:elem-disjunction} shows that we
can compute the width and maximal order type of all elementary \wqos
in normal form, as summarized in \Cref{tab:fam:w-o-compute}.

\begin{table}[ht]
    \centering
    \begin{tabular}{c c c}
        \toprule
        \textbf{E}  & $\w(E)$ & $\o(E)$  \\
        \midrule
        $\alpha,E_1^\seq, E_1 \sqcup E_2$  & \multirow{2}{*}{\Cref{tab:fam:la-famille}}& \multirow{2}{*}{\Cref{tab:fam:la-famille}} \\
        $\Mul(E_1) $ & &
        \\
        \midrule
        $\Pf(E_1)$	& $2^{\o(E_1)}$ & $2^{\o(E_1)}$\\
        \midrule
        $E_1\times E_2$	&
        $\o(E_1) \otimes \o(E_2)$ &
            $\o(E_1) \otimes \o(E_2)$\\
        \bottomrule
    \end{tabular}
      \caption{Width and m.o.t.\ for elementary wqos}%
    \label{tab:fam:w-o-compute}
\end{table}

\subsection{Height of elementary wqos}%
\label{sec:elementary-height}

Given a \wqo $E$, its height $\h(E)$ is computable from the heights of its
sub-expressions (see \Cref{tab:fam:la-famille}), with the notable exception of
$\h(\Pf(E))$ which cannot be expressed as a function of $\h(E)$ as proven in
\Cref{sec:tightness}.

\AP Recall that the height $\h(\Pfull(E))$ of the full powerset construction
can be computed via $\h(\Pfull(E)) = \o(E) + 1$ (see \Cref{rem-Pinf}). The main
contribution of this section, \Cref{lem:fam:h-pf-oinf}, leverages this equality
to connect the height of the finitary powerset to the supremum of the maximal
order type of \emph{approximations} of $E$, that we will introduce in
\Cref{def:approximation-wqo,def:approximate-invariants}.

\AP Approximations will be defined inductively on the parse tree of an
\kl{elementary \wqo}, by replacing constructors with ``strict approximations.''
For instance, an ordinal $\alpha$ will be approximated by any smaller ordinal
$\alpha'<\alpha$, and the \kl{finitary sequence} construction $E^\seq$ will be
approximated by \kl{Cartesian products} of the form $E^{\times n}$ for $n<\om$.
To approximate \kl{finite multisets}, we define for any \wqo $A$ and for any
$n<\om$, the set $\intro*\Muln(A)$ to be the set of multisets of exactly $n$
elements ordered with the \kl{multiset embedding}.

\begin{definition}[Approximation of an elementary \wqo]
    \label{def:approximation-wqo}
  Let $E,E'$ be two \kl{elementary \wqos}.
    We say $E'$ \intro{approximates} $E$, denoted by $E' \intro*\leqapprox
    E$, iff one of the following clauses holds:\footnote{The
    recursion in the definition is well-founded so the relation
    $\leqapprox$ is well defined.}
	\begin{itemize}
		\item $E=\alpha$ and $E'\wpoiso\alpha'<\alpha$,
		\item $E=E_1\sqcup E_2$ and $E'\wpoiso E_1'\sqcup E_2'$ where $E_1'\leqapprox E_1$ and $E_2'\leqapprox E_2$,
		\item $E=E_1\times E_2$ and $E'\wpoiso E_1'\times E_2'$ where $E_1'\leqapprox E_1$ and $E_2'\leqapprox E_2$,
		\item $E=E_1^\seq$ and $E'\wpoiso (E_1')^{\times n}$ where $E_1'\leqapprox E_1$ and $n<\om$,
		\item $E=\Mul(E_1)$ and $E'\wpoiso \Muln(E_1')$ where $E_1'\leqapprox E_1$ and $n<\om$,
		\item $E=\Pf(E_1)$ and $E'\wpoiso \Pf(E_1')$ where $E_1'\leqapprox E_1$.
	\end{itemize}
\end{definition}

\AP The notion of \kl{approximation} can be understood as a principled way of
considering strictly simpler substructures of a given \kl{elementary $\wqo$}
$E$, and the following fact ensures that \kl{approximations} are indeed
\kl{substructures} of $E$, proven using a straightforward induction on its
expression.

\begin{fact}
  \label{fact:approx-leqstruct}
    $E' \leqapprox E \implies E' \leqstruct E$.
\end{fact}

\begin{example}
  \label{ex:approximation}
  Let $E = (\om^\om \cartprod \Pf(\om^\om))^\seq$.
  Typical approximations of $E$ include
  the sets
  $((\om\cdot n)\cartprod \Pf(\om\cdot n))^{\cartprod n}$
  for $n < \om$.
  Indeed, those are \kl{approximations} of $E$,
  and any other \kl{approximation} $E' \leqapprox E$ is
  \kl{\wpo-isomorphic} to a \kl{substructure} of one of those.
\end{example}

The key idea behind the introduction of \kl{approximations} is to introduce
\emph{smoothed} (weakened) versions of the ordinal invariants, that will
behave better with respect to the \kl{finitary powerset} construction. These
new invariants are defined as the supremum of the corresponding invariant over
all \kl{approximations}.

\begin{definition}[Weakened ordinal invariants]
  \label{def:approximate-invariants}
  Let $\someInvariant \in \{ \o, \h, \w \}$ be an \kl{ordinal invariant}, and $E$ be an
  \kl{elementary \wqo}.
	Then, the
  \intro{weakened ordinal invariant} associated to $\someInvariant$
  is defined as
  $\intro*\someApproxInvariant(E) \defined \osup{\someInvariant(E') + 1}{E' \leqapprox E}$.
\end{definition}

Let us illustrate the computation of $\oinf(E)$ on the example from
\cref{ex:approximation}.

\begin{example}
  \label{ex:weakened-invariants}
  Let $E = (\om^\om \cartprod \Pf(\om^\om))^\seq$ as in
  \Cref{ex:approximation}.
  Then, we have
  \begin{align*}
    \oinf(E)
    &=
    \osup{
      \o\bigl(((\om \cdot n) \cartprod \Pf(\om \cdot n))^{\cartprod n}\bigr) + 1
    }{n < \om}
    \\
    &=
    \osup{
       \bigl( \o(\om\cdot n) \otimes \o(\Pf(\om \cdot n)) \bigr)^{\otimes n}
      +1
    }{n < \om}
    \\
    &=
    \osup{
      \left( (\om \cdot n) \otimes (\om\cdot n) \right)^{\otimes n} +1
    }{n < \om}
    \\
    &=
    \osup{
      \om^{2n} \cdot n^{2n}
      +1
    }{n < \om}
    \\
    &=
    \om^{\om} \,,
  \end{align*}
  where $\alpha^{\otimes n}$ denotes the natural product $\alpha\otimes \dots \otimes \alpha$ $n$ times.
  On the other hand, using the inductive definition of
  the \kl{maximal order type} from \Cref{tab:fam:w-o-compute}
  on \kl{elementary \wqos},
  we have
$\o(E) = \om^{\om^{\om^{\om\cdot 2}}}$
  which is much larger than $\oinf(E)$.
\end{example}

As for ordinary ordinal invariants, we are going to use structural monotonicity
on a regular basis, which is ensured by the following lemma that extends
\Cref{lem:invariants-monotonicity} to \kl{approximations}.

\begin{fact}
\label{fact:approx-subexpr}
Let $E_1,E_2$ be \kl{elementary \wqos} such that the normal form of $E_1$ is a sub-expression of the normal form of $E_2$,
and let $\someInvariant \in \{\o,\h,\w\}$.
Then $\someApproxInvariant(E_1)\leq \someApproxInvariant(E_2)$.
\end{fact}
\begin{proof}
If $E_1$ is a sub-expression of $E_2$, then $E_1$ is isomorphic to a substructure of $E_2$. Furthermore, for any $E'_1\leqapprox E_1$, there exists $E'_2\leqapprox E_2$ such that $E'_1\leqstruct E'_2$. Hence $\someInvariant(E'_1)\leq \someInvariant(E'_2)$ according to \cref{lem:invariants-monotonicity}, therefore $\someApproxInvariant(E_1)\leq \someApproxInvariant(E_2)$.
\end{proof}

We are now ready to state the main technical result of this section,
stating that computing $\oinf(E)$ is sufficient to compute $\h(\Pf(E))$.
\begin{restatable}{therm}{theoremApproxOinf}
	\label{lem:fam:h-pf-oinf}
	For every elementary \wqo $E$,
	    $\h(\Pf(E)) = \oinf(E)$.
\end{restatable}

From this theorem and \Cref{ex:weakened-invariants}, one concludes that the
\kl{finitary powerset} of the \kl{elementary \wqo} $E$ from
\Cref{ex:approximation} has \kl{height} $\h(\Pf(E)) = \oinf(E) = \om^{\om}$.

\paragraph*{Computing $\oinf(E)$.} Before proving \Cref{lem:fam:h-pf-oinf}, let
us first show how to compute $\oinf(E)$ for any \kl{elementary \wqo} $E$, that
will make the situation of \Cref{ex:weakened-invariants} systematic. The rules
for computing $\oinf(E)$ are summarized in \Cref{tab:fam:oinf-elem}, where some
hypotheses are required to ensure the correctness of the computations. After
establishing these rules in \Cref{lem:fam:oinf-elem}, we will show in
\Cref{lem:fam:taming} that these hypotheses are always satisfied for
\kl{elementary \wqos}.

\begin{table}[ht]
	\begin{center}
		\begin{tabular}{c c l}
			\toprule
			\textbf{$E$}     & \textbf{$\oinf(E)$}            & \textbf{Hypothesis}       \\
			\midrule
			$E_1 \cartprod E_2$ & $\max(\oinf(E_1), \oinf(E_2))$
			& $\oinf(E_1), \oinf(E_2)$ \kl{indecomposable}                          \\
			\addlinespace
			$E_1 \dunion E_2$ & $\max(\oinf(E_1), \oinf(E_2))$
			& $\oinf(E_1), \oinf(E_2)$ \kl{indecomposable}                          \\
			\addlinespace
			$\Mul(E_1)$      & $\oinf(E_1)$                                               & $\oinf(E_1)$ \kl{indecomposable}                          \\
			\addlinespace
			$E_1^\seq$          & $\oinf(E_1)$                                               & $\oinf(E_1)$ \kl{indecomposable}                          \\
			\addlinespace
			$\Pf(E_1)$       & $2^{\oinf(E_1)}$
                        & $\oinf(E_1) = \winf(E_1)$ limit \\
			\bottomrule
		\end{tabular}
	\end{center}
	\caption{Computing the weakened maximal order type under conditions.}
	\label{tab:fam:oinf-elem}
\end{table}

\begin{lemma}
  \label{lem:fam:oinf-elem}
  For $E$ \kl{elementary}, $\oinf(E)$  can be computed as described
  in \Cref{tab:fam:oinf-elem} in cases following the given hypotheses.
\end{lemma}
\begin{proof}
  Let $E_1, E_2$ be \kl{elementary \wqos}, let
	$\alpha = \oinf(E_1)$, and let $\beta = \oinf(E_2)$
\begin{itemize}
  \item Without loss of generality, let us assume that $\beta \leq \alpha$. Then, with the hypothesis that $\alpha$ and $\beta$ are \kl{multiplicatively indecomposable}:
	\begin{itemize}
    \item
		      $\oinf(E_1 \dunion E_2) \leq
			      \sup \setof{\alpha' \oplus \beta'+1}{\alpha'<\alpha,\beta'<\beta}
			      =\alpha$.
		\item
		      $\oinf(E_1 \cartprod E_2) \leq
			      \sup \setof{\alpha'\otimes\beta'+1}{\alpha'<\alpha,\beta'<\beta}
			      =\alpha$.

		\item
		      $\oinf(E_1^\seq) \leq
			      \sup \setof{(\alpha')^{\otimes n}+1}{\alpha'<\alpha,n<\om}\leq\alpha$
		\item
		      $\Muln(A)\geqaug (A)^{\times n}$ for any \wqo $A$,
		      hence
		      $\oinf(\Mul(E_1)) \leq \oinf(E_1^\seq)=\alpha$.
\end{itemize}
Observe that in these four cases, $E_1$ is a sub-expression of some
$E$ with $\alpha\leq\oinf(E)$ according to \Cref{fact:approx-subexpr}.
		\item If both $\winf(E_1)$ and $\oinf(E_1)$ are limit then $2^{\winf(E_1)} \leq \winf(\Pf(E_1))
			      \leq \oinf(\Pf(E_1)) \leq 2^{\oinf(E_1)}$ according to
			      \Cref{thm:pf-width,thm:pf-mot}.
		      Hence if also $\oinf(E_1) = \winf(E_1)$, then
		       $\oinf(\Pf(E_1)) = 2^{\oinf(E_1)}$.
		      \qedhere
	\end{itemize}
\end{proof}
The condition $\oinf(E_1)=\winf(E_1)$ for $E=\Pf(E_1)$ in
\Cref{tab:fam:oinf-elem} is not very restrictive, since operations like
$E^\seq$ and $\Mul(E)$ do not drastically increase the \kl{weakened m.o.t.} of
an elementary \wqo as they do for the \kl{m.o.t.}, to the point that
$\oinf=\winf$ holds for most elementary \wqos. To prove that, we develop some
lower bounds on $\winf$.

One of the difficulty in computing the \kl{weakened width} will come from the
\kl{finitary sequence} construction and the \kl{finite multiset} construction,
that will essentially require us to compute the \kl{width} of structures of the
form $\Muln(\alpha)$. As we will see, it suffices for our purposes to
focus on $\Multwo(\alpha)$ to obtain our desired lower bounds on $\winf$.

\begin{definition}
  \label{def:increasing-product}
  For any ordinal $\alpha$, let us define
  $\intro*\IncPairs{\alpha}
  \defined \Multwo(\alpha)$.
\end{definition}

Remark that $\IncPairs{\alpha}$ can be seen as the set of pairs of ordinals
$(\beta_0,\beta_1)$ such that $\beta_0 \leq \beta_1 < \alpha$, ordered
point-wise, and therefore corresponds to a \kl{substructure} of the
\kl{Cartesian product} $\alpha \times \alpha$. Furthermore, $\IncPairs{\alpha}$
\kl{reflects into} $\Pf(\alpha \times 2)$ through the function
$(\beta_0,\beta_1)\mapsto \{(\beta_1,0),(\beta_0,1)\}$.

Let us now establish a lower bound on $\winf(\IncPairs{\alpha})$, which is done
by extending the definition of $\winf$ to well-quasi-orders of the form
$\Multwo(\alpha)$ as follows: $E' \leqapprox \Multwo(\alpha)$ iff $E' \wpoiso
\Multwo(\alpha')$ for some $\alpha'<\alpha$. As a consequence of
this definition, we always have $\winf(\Mul(\alpha)) \geq
\winf(\IncPairs{\alpha})$.

\begin{lemma}
	\label{lem-w-produit-croissant}
	Let $\alpha\geq\omom$ be an \kl{indecomposable} ordinal. Then,
    \begin{equation*}
      \winf(\IncPairs{\alpha}) =
        \sup_{\alpha'<\alpha} \w(\IncPairs{\alpha'})+1\geq\alpha
        \:.
    \end{equation*}
\end{lemma}
\begin{proof}
For any $\om\leq\alpha'<\alpha$ and any decreasing sequence
$\alpha'-\om >\beta_0 >\beta_1>\dots$ the sequence
$((k,\beta_k+\om))_{k=0,1,\ldots}$ is an \kl{antichain} in
$\IncPairs{\alpha'}$.
This provides a map $\DecSeq(\alpha'-\om) \to
\IncSeq(\IncPairs{\alpha'})$ witnessing
$\w(\IncPairs{\alpha'})\geq\h(\alpha'-\om)=\alpha'-\om$.  Hence
$\winf(\IncPairs{\alpha}) \geq \sup_{\alpha'<\alpha}(\alpha'-\om)+1 =
\alpha - \om =\alpha$ since $\alpha$ is \kl{indecomposable}.
\end{proof}
As we did with \Cref{lem:elem-disjunction}, let us present a structural lemma
which reveals in which cases $\winf\neq\oinf$.

\begin{lemma}[Case disjunction]
	\label{lem:fam:taming}
  Let $E$ be an \kl{elementary \wqo} (given through its expression in normal form).
  Then, $\oinf(E)$ is \kl{indecomposable}, and one of
	the following holds:
	\begin{compactenum}[(i)]
		\item $\winf(E) = \oinf(E)$,
		\item $E=E_1\dunion E_2$ a disjoint union of elementary \wqos,
		\item $E=\alpha$ an ordinal.
	\end{compactenum}
\end{lemma}

\begin{proof}
  Observe that for any elementary \wqo $E$, $\winf(E)\leq\oinf(E)$ because
  $\w(E')\leq \o(E')$ for any $E'\leqapprox E$. Therefore, to conclude that
  $\oinf(E) = \winf(E)$, it suffices to prove that $\winf(E) \geq \oinf(E)$.
  Let us now proceed by induction on the expression of $E$, but before that,
  let us point out that aside from $\Pf$, the \kl{weakened maximal order type}
  of every elementary operation can be computed under an indecomposability
  hypothesis. Moreover \Cref{tab:fam:oinf-elem} shows how the indecomposability
  of the \kl{weakened m.o.t.} is preserved by our constructions.

\begin{itemize}
\item [Case $E = \alpha$ \kl{indecomposable}] satisfies $(iii)$, and we
  can check that $\oinf(\alpha)=\alpha$ is \kl{indecomposable}.

\item [Case $E=\Pf(E_1)$:] By induction hypothesis $E_1$ verifies $(i)$ since $E$ is in normal form, hence cannot be an ordinal nor a disjoint union from the rules of \Cref{fig:rewrite-rules}.
        Therefore
        $\winf(E) = \oinf(E)=2^{\oinf(E_1)}$
        applying \Cref{tab:fam:oinf-elem}.
        Hence $E$ verifies $(i)$.
        Finally, because $\oinf(E_1)$ is \kl{indecomposable} according to the induction hypothesis,
        we conclude that $\oinf(E)=2^{\oinf(E_1)}$ is \kl{indecomposable} as well.

        \item [Case $E = E_1 \sqcup E_2$] satisfies $(ii)$.

        \item [Case $E = E_1 \times E_2$:] Assume that
        $\oinf(E_1) = \alpha \geq \beta = \oinf(E_2)$,
        and therefore that $\oinf(E)=\alpha$ (see \Cref{tab:fam:oinf-elem}).
        Now there exists
        $E_2'\leqapprox E_2$ such that $\om \leqstruct E_2'$.
        Thus for any $E_1'\leqapprox E_1$,
        one can leverage the structural monotonicity of the width
        (\Cref{lem:invariants-monotonicity}) and the specific
        inequality between \kl{width} and \kl{m.o.t.} of
        \Cref{lem-isa} to obtain:
        \begin{align*}
          \w(E_1' \times E_2')
          \geq \w(E_1' \times \om)
          \geq \o(E_1')
        \end{align*}
        We conclude that $\winf(E) \geq \oinf(E_1) = \oinf(E)$,
        proving that
        $E$ verifies $(i)$.

        \item [Case $E = \Mul(E_1)$:]
            For all $E_1'\leqapprox E_1$, we know that
            $\Multwo(E_1')\leqapprox E$ by definition
            of \kl{approximation}.
            Since $E_1' \leqaug \o(E_1')$,
            and because \kl{finite multisets} are monotonic with respect to
            \kl{augmentations},
            we have
            $\Multwo(E_1')\leqaug \Multwo(\o(E_1')) = \IncPairs{\o(E_1')}$.
            As a consequence, and according to
            \Cref{lem-w-produit-croissant}:
            \begin{align*}
              \winf(E) &\geq
              \osup{ \w(\IncPairs{\o(E_1')})}{E_1' \leqapprox E_1} \\
                       &\geq\winf(\IncPairs{\oinf(E_1)}) \\
                       &\geq \oinf(E_1)
            \end{align*}
          Furthermore,
          $\oinf(E) = \oinf(E_1)$ according to \Cref{tab:fam:oinf-elem},
          and we have proven that $E$ satisfies $(i)$.

        \item [Case $E = E_1^\seq$:]
 For any $E_1'\leqapprox E_1$, for any $n<\om$, $\Muln(E_1')\geqaug (E_1')^{\times n}$, thus $\winf(E)\geq \winf(\Mul(E_1))$.
 Since $\oinf(E) = \oinf(E_1) = \oinf(\Mul(E_1))$ according to \Cref{tab:fam:oinf-elem}, and by the previous case, we have $\winf(E)\geq \oinf(E)$.
 Hence $E$ verifies $(i)$.

		\qedhere
    \end{itemize}
\end{proof}

Hence, the hypotheses in \Cref{tab:fam:oinf-elem} are verified for all
elementary \wqo $E$, and assuming that $E$ is in normal form, we can
inductively compute $\oinf(E)$ without any hypothesis, as summarized in
\Cref{tab:fam:h-pf-computation}.

\begin{table}[ht]
	\begin{center}
		\begin{tabular}{cc}
			\toprule
			\textbf{$E$ in normal form}     & \textbf{$\h(\Pf(E))=\oinf(E)$}                      \\
			\midrule
			$\alpha$ & $\alpha$\\
			$E_1 \times E_2$ & $\max(\oinf(E_1),\oinf(E_2))$                             \\
			\addlinespace
			$\Mul(E_1)$      & $\oinf(E_1)$                                                 \\
			\addlinespace
			$E_1^\seq$          & $\oinf(E_1)$                                                 \\
			\addlinespace
			$\Pf(E_1)$
			                 & $2^{\oinf(E_1)}$                                     \\
			\bottomrule
		\end{tabular}
	\end{center}
	\caption{Computing the weakened maximal order type inductively.}
	\label{tab:fam:h-pf-computation}
\end{table}

\begin{therm}
 $\h(E)$ is computable for elementary $E$  given by its elementary expression.
\end{therm}
\begin{proof}
	Assume $E$ is in normal form, and see
        \Cref{tab:fam:h-pf-computation,tab:fam:la-famille}.
\end{proof}

\paragraph*{Proving \Cref{lem:fam:h-pf-oinf}.}
It remains for us to prove the main theorem of this section, that we restate
hereafter for readability.

\theoremApproxOinf*

\AP The easiest inequality to obtain is the lower bound $\oinf(E) \leq
\h(\Pf(E))$. It can be dealt with through properties of ideal completions.
Recall that an \intro{ideal} $I$ of a \wqo $A$ is a non-empty \kl{downwards
closed} and \kl{up-directed} subset of $A$, see, e.g.,~\cite{ghkks-ideals}. For
completeness, let us recall that a set $I$ being \intro{up-directed} means that
for every $a, b \in I$, there exists $c \in I$ such that $a \leq c$ and $b \leq
c$.

\begin{definition}[Ideal Completion]
  The set of \kl{ideals} of $A$ ordered with \kl{Hoare's embedding} is denoted
$\intro*\Idl(A)$, and is called the \intro{ideal completion} of $A$.\footnote{
When $A$ is a \wpo,   ordering ideals by inclusion or by
Hoare's embedding is equivalent.
}
\end{definition}

The goal of this article is not to study ideal completions in depth,
and we will only use some basic properties that were already established
in prior work~\cite{halfon2018}.

\begin{fact}[See \cite{halfon2018}]
	\label{fact-idl-com}
	\label{fact-idl-down-h}
  The following properties hold for any \wqo $A,B$ and ordinal $\alpha$:
\begin{compactenum}[(1)]
\item  Ideal completion commutes with
          $\sqcup$, $\times$, $\Muln$ and $\Pf$, i.e., $\Idl(A\sqcup
          B)\cong\Idl(A)\sqcup\Idl(B)$, etc.,
\item $\Idl(\alpha) \cong \alpha + 1$,
\item $\Pf(\Idl(A))\cong \Pfull(A)$.
\end{compactenum}
\end{fact}

With point (3) above we see  that $\Idl(A)$ is not always a \wqo, as
$\Pfull(A)$ may not be one! However, since an \kl{approximation} $E'$ of an
elementary $E$ only uses constructions that commute with \kl{ideal completion},
we get \Cref{fact:idl-approx}. This fact not only states that the \kl{ideal
completion} of an \kl{approximation} is again a \wqo, but that it even remains
an \kl{approximation} of the original \kl{elementary \wqo}. This will be
particularly useful, because $\h(\Pfull(E)) = \o(E) + 1$ for any \wqo $E$, as
we saw in \Cref{rem-Pinf}.

\begin{fact}
    \label{fact:idl-approx}
Assume $E' \leqapprox E$ for an elementary  \wqo $E$.
    Then $\Idl(E') \leqapprox E$.
\end{fact}

\begin{lemma}
	\label{lem:fam:pf-oinf-down}
	For every elementary \wqo $E$, $\oinf(E) \leq \h(\Pf(E))$.
\end{lemma}
\begin{proof}
	Let $E' \leqapprox E$. Then $\Idl(E') \leqapprox E$
    thanks to \Cref{fact:idl-approx}, and
    in particular $\Idl(E') \leqstruct E$ (\Cref{fact:approx-leqstruct}), thus $\Pf(\Idl(E')) \leqstruct \Pf(E)$.

    Using the monotonicity of the \kl{height} with respect to
    \kl{substructures} (\Cref{lem:invariants-monotonicity}), we conclude that
    $\h(\Pf(E)) \geq \h(\Pf(\Idl(E')))$. Now, by \Cref{fact-idl-com}, we know
    that $\Pf(\Idl(E')) \cong \Pfull(E')$.
    Finally, leveraging \Cref{rem-Pinf}, we obtain
    $\h(\Pf(\Idl(E'))) = \h(\Pfull(E')) = \o(E') + 1$.
    Hence, we have proven that
    $ \h(\Pf(E))\geq\oinf(E)$.
\end{proof}

The upper bound $ \h(\Pf(E))\leq\oinf(E)$, in contrast, is quite
subtle to prove.
Observe that
\emph{augmentations} (\Cref{def:augmentation})
do not preserve the height,
depriving us of one of our favorite ways to prove bounds.
That is why we introduce the notion of \emph{condensation}.

\begin{definition}[Condensation]
    \label{def:condensation}
	A mapping $f \colon A \to B$ between two \wqos is a \intro{condensation}
	if it is surjective, monotonic,
	and whenever $b \leq_B f(y)$, there exists $x \leq_A y$
	such that $b = f(x)$.
	When there exists a condensation from $A$ to $B$,
	we note $B \intro*\leqcm A$.
\end{definition}

It appears that \kl{condensations} are also known in the literature as
\emph{p-morphisms} in the context of Kripke frames and modal logic. In the
specific context of quasi-orders, it seems that it is less standard than
\kl{reflection}, \kl{augmentation}, and \kl{substructure}. Intuitively, a
condensation can be seen as a quotient, where the quotient ordering and the
original ordering are related.

\begin{example}
  The function $\iota \colon A^{\times n} \to \Muln(A)$
	mapping $(a_1, \dots, a_n)$ to the multiset
  $\mset{ a_1, \dots, a_n}$ is a \kl{condensation}.
\end{example}

\begin{remark}
  A surjective monotonic function $f \colon A \to B$ is a \kl{condensation} if and only
  if the image of any \kl{downwards-closed}
  set is itself \kl{downwards-closed}. In topological terms,
	$f$ is continuous, closed, and surjective.
\end{remark}

The key property of \kl{condensations} from $A$ to $B$ is that one can simulate
decreasing sequences $(b_i)_{i \in \Nat}$ occurring in $B$ through a careful
selection of pre-images $a_i \in f^{-1}(b_i)$. As a consequence, the
\kl{height} of $B$ is bounded by the \kl{height} of $A$.

\begin{lemma}
  If $B \leqcm A$, then $\h(B) \leq \h(A)$.
\end{lemma}
\begin{proof}
  Let $f \colon A \to B$ be a \kl{condensation}, $b_1>\dots> b_{n+1}$
	a strictly decreasing sequence in $B$, and $a_1>\dots>a_n$ a strictly
	decreasing sequence in $A$ such that $f(a_i)=b_i$ for all $i\leq n$.

  Since $b_{n+1}\leq b_n$ there exists $a_{n+1}\leq a_n$ such that
  $f(a_{n+1})=b_{n+1}$. Since $f$ is monotonic, $b_{n+1} \not\geq b_n$ implies
  that $a_{n+1}\not\geq a_n$. Hence $a_1>\dots>a_{n+1}$. This shows that with
  any strictly decreasing sequence of $B$ we can associate a strictly
  decreasing sequence of $A$ in a way that respects prefix order, hence
  $\DecSeq(B)$ is a substructure of $\DecSeq(A)$ modulo isomorphism.
  We have proven that $\h(B) \leq \h(A)$.
\end{proof}

\begin{example}
	The function $\iota \colon \Mul(A) \to \Pf(A)$
	mapping a multiset $M$ to the set
	$\setof{a}{a\in M}$ is monotonic, surjective, but is
	not always a condensation.
\end{example}
\begin{proof}
Take $A=3$. In $\Pf(A)$ one has $\{0,1\}\hoareleq\{2\}$, but
no multiset $M$ in $\iota^{-1}(\{0,1\})$ is dominated by $\mset{2}$ since
any such $M$ contains at least
one $0$ and one $1$.
More generally, we know of instances where $\h(\Mul(A))<\h(\Pf(A))$ (see family $\FSim{\alpha}$ in \Cref{sec-palpha}).
\end{proof}

We now establish two properties of \kl{condensations}.  First we show that the
constructors used to build \kl{approximated \wqos} are monotonic with respect
to $\leqcm$ (\Cref{fact:leqcm-monotone}). Then we show that approximations can
be extended while respecting $\leqcm$ (\Cref{lem:one-plus-approx-is-approx}).
We will actually consider a variation $\leqcmstruct$ of $\leqcm$, that we
formally state hereafter.

\begin{definition}
    \label{def:leqcmstruct}
    We define $\intro*\leqcmstruct$ as the transitive closure of
    the relations $\leqcm$ and $\leqstruct$, which is a
    quasi-ordering over \wqos.
\end{definition}

\begin{fact}[Monotonicity]
    \label{fact:leqcm-monotone}
	The operations $\sqcup$, $\times$, $\Pf$, $\Mul$, $^{<\om}$ and $+$ are monotonic with respect to $\leqcm$.
\end{fact}

\begin{lemma}
\label{lem:one-plus-approx-is-approx}
Let $E$ be an elementary \wqo and $E'\leqapprox E$. Then there exists $E''\leqapprox E$ such that $1 \lexsum E'\leqcmstruct E''$.
\end{lemma}
\begin{proof}
By induction on the expression of $E$ in normal form. First observe that:
\begin{itemize}
  \item [Case $E=\alpha$:]
    for all $\alpha'<\alpha$, $1+\alpha'<\alpha$ since $\alpha$ is infinite.

    \item [Case $E=E_1\dunion E_2$:] $1 \lexsum (E_1'\dunion E_2') \leqcm (1
      \lexsum E_1')\dunion (1+E_2')$ for all $E_1'\leqapprox E_1$,
      $E_2'\leqapprox E_2$. Here the \kl{condensation} function maps the two
      minimal elements of $(1 \lexsum E_1')\dunion (1\lexsum E_2')$ to the
      unique minimal element of $1+(E_1'\dunion E_2')$.

		\item [Case $E=E_1\cartprod E_2$:] $1\lexsum (E_1'\cartprod E_2')
      \leqstruct (1 \lexsum E_1')\cartprod  (1 \lexsum E_2')$.

    \item [Case $E=E_1^\seq$:] $1 \lexsum (E_1')^{\times n}\leqstruct (1 \lexsum E_1')^{\times n}$.

		\item [Case $E=\Mul(E_1)$:] $1 \lexsum \Muln(E_1')\leqstruct \Muln(1 \lexsum E_1')$.

		\item [Case $E=\Pf(E_1)$:] $1 \lexsum \Pf(E_1')\leqstruct \Pf(1 \lexsum E_1')$.
\end{itemize}

By induction hypothesis, for any $E_i'$  with $i\in\{1,2\}$ appearing in the
cases above, there exists $E_i''\leqapprox E_i$ such that $1\lexsum E_i'\leqcmstruct
E_i''$. We conclude using \cref{fact:leqcm-monotone}.
\end{proof}

\begin{lemma}
	\label{lem:fam:over-approx-pf}
  Let $E$ be an \kl{elementary \wqo}. For all $S \in \Pf(E)$, there exists $E'\leqapprox E$ such that $\dwc_E S \leqcmstruct E'$.
\end{lemma}
\begin{proof}
	By induction on the expression in normal form of $E$:
	\begin{itemize}
		\item[Case $E=\alpha$:] Let $\gamma = \max\set{x\in S}$.
		      Then $\dwc S \cong \gamma \leqapprox E$.

		\item[Case $E=E_1\dunion E_2$:]
		$\dwc_E S \cong \dwc_{E_1} S_1 \sqcup \dwc_{E_2} S_2$
		      where $S_1 = S \cap E_1$ and $S_2 = S \cap E_2$.
		      By induction hypothesis there exist $E_1',E_2'\leqapprox E_1,E_2$ such that
		      $\dwc_{E_1} S_1 \leqcmstruct E_1'$, $\dwc_{E_2} S_2 \leqcmstruct E_2'$.
		      Hence,
              \begin{equation*}
                  \dwc_E S \leqcmstruct E_1' \sqcup E_2'
                  \:.
              \end{equation*}

		\item[Case $E=E_1\cartprod E_2$:] $\dwc_E S \leqstruct \dwc_{E_1} S_{|E_1} \times \dwc_{E_2} S_{|E_2}$.
		      By induction hypothesis there exist $E_1',E_2'\leqapprox E_1,E_2$ such that
		      $\dwc_{E_1} S_{|E_1}\leqcmstruct E_1'$ and $\dwc_{E_2} S_{|E_2} \leqcmstruct E_2'$.
		      Hence,
              \begin{equation*}
                  \dwc_E S \leqcmstruct E_1' \cartprod E_2'
                  \:.
              \end{equation*}

		\item[Case $E=E_1^\seq$:]
		      Let $n$ be the maximal length of words in $S$,
		      and $S'\in\Pf(E_1)$ the set of letters in words of $S$.
		      Then $\dwc_E S \leqstruct (\dwc_{E_1} S')^{\leq n}$.
		      Observe that $(\dwc_{E_1} S')^{\leq n}\leqcm (1 \lexsum \dwc_{E_1} S')^{\times n}$:
          indeed elements of $(1 \lexsum \dwc_{E_1} S')^{\times n}$
          can be seen as words in $(\dwc_{E_1} S')^{\leq n}$
          padded with extra bottom elements, with the \kl{condensation} function removing the padding.
		      By induction hypothesis, there exists $E_1'\leqapprox E_1$ such that $\dwc_{E_1}
			      S'\leqcmstruct E_1'$.

              Thanks to
              \Cref{lem:one-plus-approx-is-approx},
              there exists
              some $E_1''\leqapprox E_1$
              such that
              \begin{equation*}
                  \dwc_{E} S \leqcmstruct (1 \lexsum  E_1')^{\times n}\leqcmstruct
              (E_1'')^{\times n}
                  \:.
              \end{equation*}

		\item[Case $E=\Mul(E_1)$:] Let $n$ be the maximal cardinal of multisets in $S$,
		      and $S'\in\Pf(E_1)$ the set of elements in multisets of $S$.
		      Then $\dwc_E S \leqstruct \Mul_{\leq n}(\dwc_{E_1} S')\cong \Muln(1 \lexsum \dwc_{E_1} S')$.
		      By induction hypothesis, there exists some $E_1'\leqapprox E_1$ such that
              $\dwc_{E_1} S'\leqcmstruct E_1'$.

			  Therefore,
              there exists
              some $E_1''\leqapprox E_1$ given by
              \Cref{lem:one-plus-approx-is-approx}, such that
              \begin{equation*}
                  \dwc_{E} S \leqcmstruct \Muln(1 \lexsum E_1')\leqcmstruct \Muln(E_1'')
                  \:.
                \end{equation*}
		\item[Case $E = \Pf(E_1)$:] Let $S'=\setof{x \in E_1}{\exists y \in S \text{ s.t. } x\in y}
		\in\Pf(E_1)$. Then $\dwc_{E} S \leqstruct \Pf(\dwc_{E_1} S')$.
		Using the induction hypothesis, there exists $E_1'\leqapprox E_1$ such that $\dwc_{E_1} S' \leqcmstruct E_1'$, hence
            \begin{equation*}
                \dwc_{E} S \leqstruct \Pf(E_1')
                \:. \qedhere
            \end{equation*}
	\end{itemize}
\end{proof}

\begin{lemma}
	\label{lem:fam:pf-oinf-up}
	$\h(\Pf(E)) \leq \oinf(E)$.
\end{lemma}
\begin{proof}
	Recall that, by Eq.~\eqref{eq-Des-h},
    \begin{equation*}
	    \h(\Pf(E)) = \sup_{S\in\Pf(E)}\h(\Pf(E)_{< S}) + 1 \:.
    \end{equation*}
	Notice that, given $S \in \Pf(E)$, we have
    $\h(\Pf(E)_{< S}) + 1
		= \h(\Pf(E)_{\leq S})$ and $\Pf(E)_{\leq S}\leqstruct \Pf(\dwc_E S)$.

	Using \Cref{lem:fam:over-approx-pf},
	there exists $E' \leqapprox E$ such that
	$\dwc_E S \leqcmstruct E'$.
	As a consequence,
	$\h(\Pf(E)_{\leq S}) \leq \h(\Pf(E')) \leq \h(\Pfull(E'))\leq \o(E') + 1$ according to \Cref{fact-idl-down-h}.

	Therefore, for every $S \in \Pf(E)$,
	$\h(\Pf(E)_{< S}) + 1 \leq \oinf(E)$, which implies by Eq.~\eqref{eq-Des-h}
	that
	$\h(\Pf(E)) \leq \oinf(E)$.
\end{proof}

Combining \Cref{lem:fam:pf-oinf-up,lem:fam:pf-oinf-down} proves
\Cref{lem:fam:h-pf-oinf}: $\h(\Pf(E)) = \oinf(E)$.

\subsection{Summary}

\Cref{tab:computing-elementary} summarizes how to compute the \kl{ordinal
invariants} of \kl{elementary \wqos} expressed in normal form. We redirect the
reader to \Cref{tab:fam:h-pf-computation} when it comes to computing
$\oinf(E_1)$.

\begin{table}[ht]
	\begin{center}
		\renewcommand{\arraystretch}{1.5}
		\begin{tabular}{c|c c c}
			\toprule
			 $E$            & $\h(E)$     & $\o(E)$ &
			$\w(E)$
            \\
			\midrule
			$\alpha,E_1\sqcup E_2$,  && \multirow{2}{*}{see \Cref{tab:fam:la-famille}}&\\
			$\Mul(E_1)$, $E_1^\seq$ &&&\\
			\midrule
			\addlinespace
            $E_1 \times E_2$
						& $\h(E_1)\hplus \h(E_2)$
						& $\o(E_1)\otimes\o(E_2)$
                        & $\o(E_1)\otimes\o(E_2)$ \\
			\addlinespace
			\midrule
            $\Pf(E_1)$
                            & $\oinf(E_1)$
                            & $2^{\o(E_1)}$
                            & $2^{\o(E_1)}$
            \\
			\bottomrule
		\end{tabular}
	\end{center}
	\caption{How to compute ordinal invariants of elementary wqos. %
                We assume that $E$ is expressed in normal form.}
	\label{tab:computing-elementary}
\end{table}

\section{Conclusion}

We studied the ordinal invariants of the finitary powerset of a \wqo. We showed
these invariants could not be expressed as functions in the ordinal invariants
of the underlying \wqo, and provided tight monomorphic bounds.

Even though these upper and lower bounds are often different, in \Cref{sec:algebra-well-behaved}, we managed to
 measure these invariants exactly by restricting
ourselves to a well-behaved family of \wqos obtained through classical operations.
This part relies on new tools, e.g., we introduced weakened variations of the usual ordinal invariants, which in turn, have now opened many new questions.

Let us now propose a few avenues of research in continuity of this work.

\paragraph*{Invariants for non-wqos.} %
Width and height can more generally
 be defined over spaces that are not \wqos.
On these spaces, the bounds we established do not hold in general. For instance
the height of  $\Pf(A)$ when $A$ is well-founded can reach
the $\omega^{\h(A)}$ upper bound.

\paragraph*{Intrinsic Weakened Invariants.}
The current definition of $\oinf$, $\winf$, and $\hinf$ rely on the
definition of \emph{approximation} that is defined inductively
on the syntactic construction of the elementary \wqo. Could we generalize
the notion of approximation to arbitrary \wqos? Do the weakened ordinal invariants
allow us to recover functionality even outside of elementary \wqos?

\paragraph*{Elementary wqos.}%
In what directions can we extend our family of elementary \wqos while
maintaining the computability of the ordinal invariants?

\bibliographystyle{msclike}
\bibliography{BIBLIO/biblio} \nocite{dzamonja2020,ASS-draft}%

\appendix

\section{Appendix to \Cref{sec:up-lower-bounds} ``Upper and Lower Bounds''}

We provide the proofs missing in \Cref{sec:up-lower-bounds}.

\subsection{Structural lemmas}
\label{app:proof:mot-sum}

\motsum*

\Cref{lem:mot-sum} generalizes Thm.~3.2 from \cite{dejongh77}
where the case $\beta=1$ is handled.
\begin{proof}
	Let A be a \wqo such that $\o(A) = \alpha + \beta$. Then, by Theorem 2.13 of \cite{dejongh77}, there exists a surjective reflection $f: \alpha+\beta \rightarrow A$.
	We let $A_\alpha = f(\{\gamma~|~\gamma<\alpha\})$
        and $A_\beta = f(\{\gamma~|~\alpha\leq\gamma<\alpha+\beta\})$.
	We further assume that $f$ makes $\o(A_\alpha)$ minimal: since the
	collection of all reflections from $\alpha+\beta$ to $A$ is a set, the collection of
	all possible $\o(A_\alpha)$ is a set of ordinals, which admits a
	minimal element.

	Now, since $f$ is a reflection, for all $a\in A_\alpha, b\in A_\beta$, $a\not\geq b$. It follows that
	\[
                A_\alpha \sqcup A_\beta
                \leqaug A \leqaug
		A_\alpha + A_\beta
	\]
	which implies:
	$\o(A_\alpha) + \o(A_\beta) \le \alpha + \beta \le \o(A_\alpha) \oplus \o(A_\beta)$.
	Besides, $f$ gives a reflection from $\alpha$ to $A_\alpha$, and one from $\{\gamma~|~\alpha\leq\gamma<\alpha+\beta\}$ (which is isomorphic to $\beta$ by left-cancellativity) to $A_\beta$, hence
	$\o(A_\alpha) \ge \alpha$ and $\o(A_\beta) \ge \beta$.
	But then
	$\alpha + \beta \ge \o(A_\alpha) + \o(A_\beta) \ge \alpha + \o(A_\beta)$
	and therefore $\beta \geq \o(A_\beta)$ since  ordinal addition is left-cancellative.
	Besides, since $\o(A_\alpha) \ge \alpha$, we can write
	$\o(A_\alpha) = \alpha + \gamma$ for some $\gamma$ that must satisfy
	$\gamma + \beta = \beta$.

	Then there exists a reflection $g:\alpha+\gamma \rightarrow A_\alpha$.
	We consider the reflection
	$f':\alpha + \gamma + \beta \rightarrow A$ obtained by concatenating $g$ and $f$ restricted to $\{\gamma~|~\alpha\leq\gamma<\alpha+\beta\}$.
	Let $B_\alpha \defined f'(\{\delta~|~\delta<\alpha\})$. By construction $\o(B_\alpha)\geq \alpha$.
	However, if $\gamma > 0$, then $B_\alpha$ is a $\not\geq$-residual of
	$A_\alpha$, hence $\o(B_\alpha)<\o(A_\alpha)$. It follows that
        $\o(A_\alpha)$ is not minimal, which is absurd.
	Therefore, $\gamma=0$ and $\o(A_\alpha) = \alpha$.
\end{proof}

\subsection{Lower bounds on the width of a product}

\begin{definition}[\protect{\citet{dzamonja2020}}]
	A \wqo $A$ is \emph{transferable} if
	for every $x_1, \dots, x_n \in A$,
	$\w(A_{\not\leq \{ x_1, \dots, x_n \}}) = \w(A)$.
\end{definition}

\begin{lemma}
\label{lem:h-finite}
For any \wpo $A$, if $\h(A)<\om$ then $A$ is finite.
\end{lemma}
\begin{proof}
Let $\h(A)=k<\om$. Then one can sort every element of $A$ into $k$ finite antichains, depending on their rank in $\Dec(A)$. Since every antichain of  a \wqo is finite, $A$ is finite.
\end{proof}

\begin{proposition}
	\label{prop-indecomp-transf2}
	If $\w(A) \ge \omega$ is \kl{additively indecomposable}, then $A$ contains a substructure of width
	$\w(A)$ which is transferable.
\end{proposition}

\begin{proof}
In this proof we consider $A$ as a \wpo (by quotienting by $\equiv$ if needed).
Note that for any partition $A = A_1 \uplus A_2$, $\w(A)\leq
\w(A_1)\oplus \w(A_2)$. Thus if $\w(A) \ge \omega$ is additively
indecomposable and $A_2$ is finite, necessarily $\w(A_1)=\w(A)$. In
other words, one can remove any finite number of elements from $A$
without changing its width.

Since $\w(A) \ge \om$, $A$ is infinite, hence
	$\h(A) \ge \omega$ according to \cref{lem:h-finite}.

Observe that $A_{\not\leq \{ x_1, \dots, x_n \}}= A\setminus \dwc \{ x_1, \dots, x_n \}$ for any $x_1, \dots, x_n \in A$.

	 We now proceed to prove the proposition by induction on $\h(A)$.

	If $\h(A) = \omega$ then for every $x_1, \dots, x_n \in A$,
	for $A_1 = \dwc \{ x_1, \dots, x_n \}$, we have $\h(A_1)<\h(A)=\om$ hence $A_1$ is finite according to \cref{lem:h-finite},  and therefore of finite
	width. It follows that $A_2 \defined A \setminus A_1$ is of width
	$\w(A)$ since the latter is \kl{additively indecomposable} and
	$\w(A) \le \w(A_1) \oplus \w(A_2)$. Thus $A$ is transferable.

	If $\h(A) > \omega$, either $A$ is transferable, in which case the
	proposition is proved, or it is not.

	If $A$ is not transferable then there exists $x_1, \dots, x_n \in A$ such
	that $\w(A_2) < \w(A)$ where $A_2 \defined A \setminus A_1$ and
	$A_1 \defined \dwc \{ x_1, \dots, x_n \}$.
	Since $\w(A)$ is \kl{additively indecomposable} and $\w(A) \le \w(A_1) \oplus \w(A_2)$,
	it follows that $\w(A_1) = \w(A)$. Moreover, since $\w(A)\geq\om$ is \kl{additively indecomposable}, then $\w(A'_1)=\w(A)$ for $A'_1 \eqdef A_{< x_1,\dots, x_n}$. Furthermore, $\h(A'_1) < \h(A)$ by the descent equation, while $\h(A'_1)\geq \om$ because $\w(A'_1)\geq\om$. Therefore by induction $A'_1$ has a transferable substructure of
	width $\w(A'_1) = \w(A)$.
\end{proof}

Recall the following property of transferable \wqos:
\begin{lemma}[\protect{\citet[Thm.~4.16]{dzamonja2020}}]
\label{lem-biftransferable-dss}
	If $\delta$ is an ordinal and $B$ a transferable \wqo, then
	$\w(\delta \times B) \ge \w(B) \cdot \delta$.
\end{lemma}
This lemma can be generalized to any \wqo $A$ such that $\o(A)=\delta$.
\begin{lemma}
	\label{lem:fam:bi-transferable-product}
	\label{lem-biftransferable}
	Suppose that $B$ is a transferable \wqo and $A$ any \wqo. Then $\w(A\times B)\geq \w(B)\cdot\o(A)$.
\end{lemma}
\begin{proof}
	 Since $A\leqaug \o(A)$, then $\w(A\times B)\geq
         \w(\o(A)\times B)$ and
         now \Cref{lem-biftransferable-dss} applies.
\end{proof}

\lembiftransferable*
\begin{proof}
	\label{app:proof:lem-isa}
	Since $\w(B)$ is \kl{indecomposable}, \Cref{prop-indecomp-transf2} shows
that
it has a substructure $C$ which is transferable and has width
	$\w(B)$.

	Hence $A\times B \geqstruct A \times C$ and $\w(A\times B)\geq \w(A\times C) \geq \w(B)\cdot \o(A)$ with \Cref{lem-biftransferable}.
\end{proof}

\section{Appendix to \Cref{sec:tightness} ``Tightness of the Bounds''}

\subsection{Proofs for \Cref{sec-halpha} studying the family $\FPhi{\alpha}$}

\begin{remark}\label{fact:hprod-product}

$\alpha\odot\beta$ (introduced in \cref{fig:fam:ordinal-operations}) is the \emph{Jacobsthal multiplication}. It is a transfinite iteration of the natural sum, in the same way the ordinal product is a transfinite iteration of the ordinal sum, thus it is halfway between the ordinal and natural multiplication: $\alpha\cdot\beta\leq\alpha\odot\beta\leq \alpha\otimes\beta$ (Proposition 5.1.2 of \cite{altman2017}) for any ordinals $\alpha,\beta$.

In the particular case where $\alpha$ is an additively indecomposable ordinal, we have the equality  $\alpha\cdot \beta=\alpha\odot\beta$, since $\alpha\oplus\alpha=\alpha+\alpha$.
\end{remark}

\begin{restatable}{lemma}{sumequalssupordinals}
	\label{lem:sum-equals-sup-ordinals}
	For every ordinal $\alpha > 0$,
	\begin{equation*}
		\sum_{\beta < \alpha} \omega^{\beta}
		= \begin{cases}
                        \om^{\alpha} &\text{ when $\alpha$ is a limit,}
                        \\
			\om^{\gamma} \cdot 2 & \text{ when $\alpha = \gamma + 1$ with $\gamma$ a limit,}                                           \\
			\om^{\alpha - 1}         & \text{ otherwise.}
		\end{cases}
	\end{equation*}
\end{restatable}

\begin{proof}
       \label{proof:lem:sum-equals-sup-ordinals}
       We  prove the result by induction on $\alpha > 0$.
       \begin{itemize}
               \item[For $\alpha = 1$] we have $1 = \om^{\alpha-1}$.

               \item[For $\alpha = \gamma+1$ with $\gamma$ a limit ordinal,] we have:
                     \begin{equation*}
                             \sum_{\beta < \alpha} \om^{\beta}
                             = \left(\sum_{\beta < \gamma} \om^{\beta}\right) + \om^\gamma =
                             \omega^\gamma +
                             \omega^\gamma
                             = \om^{\gamma} \cdot 2 \, .
                     \end{equation*}
               \item[For $\alpha = \gamma+2$] we have:
                     \begin{align*}
                             \sum_{\beta < \alpha} \om^{\beta}
                              & = \sum_{\beta < \gamma + 1} \om^{\beta} + \om^{\gamma+1} =
                             \begin{cases}
                                     \om^{\gamma} \cdot 2 + \om^{\gamma + 1} & \text{ when $\gamma$ limit } \\
                                     \om^{\gamma}  + \om^{\gamma + 1}        & \text{ otherwise }
                             \end{cases}                                    \\
                              & = \om^{\gamma + 1} = \om^{\alpha - 1} \, .
                     \end{align*}

               \item[For $\alpha$ a limit ordinal,]
                     $\sum_{\beta < \alpha} \om^{\beta}
                             = \osup{\sum_{\beta < \gamma} \om^{\beta}}{\gamma < \alpha}$.
                     Let $\gamma < \alpha$, then $\gamma + 2 < \alpha$
                     and
                     $\sum_{\beta < \gamma} \om^{\beta} \leq
                             \sum_{\beta < \gamma + 2} \om^{\beta}$.
                     By induction hypothesis,
                     we conclude that
                     $\sum_{\beta < \gamma} \om^{\beta} \leq \om^{\gamma+1} <
                             \om^{\alpha - 1}=\om^{\alpha}$.
                     As a consequence,
                     $\sum_{\beta < \alpha} \om^{\beta}
                             \leq \om^{\alpha}$.
                     Conversely, for every $\gamma < \alpha$,
                     $\om^\gamma \leq \sum_{\beta < \gamma+1} \om^{\beta}$,
                     and by definition of ordinal exponentiation,
                     $\om^\alpha \leq \sum_{\beta < \alpha} \om^{\beta}$. \qedhere
       \end{itemize}
\end{proof}

We are now ready to prove the main lemma.
\canofam*
\begin{proof}
\label{app:proof:prop:cano-fam}
	We prove the result by induction on $\alpha$.

	\begin{itemize}
		\item It's immediate for $\FPhi{\om^0}$.
		\item Given $x \in \FPhi{\om^1}$, $(\FPhi{\om^1})_{\not\geq x}$ is a finite set, hence $\o(\FPhi{\om^1}) \le \om$ according to Eq.~\eqref{eq-Des-o}. Since $\FPhi{\om^1}$ is an infinite sum indexed by $\om$, $\om \le \h(\FPhi{\om^1})$. Besides, $\FPhi{\om^1}$ contains arbitrarily
		      large antichains and thus $\w(\FPhi{\om^1}) \ge \om$.
		       It follows from \Cref{coro:kt} that $\h(\FPhi{\om^1}) = \w(\FPhi{\om^1}) = \o(\FPhi{\om^1}) = \om^1$.

		\item If $\alpha >0$, \Cref{tab:fam:la-famille} and the induction hypothesis give:
		      \begin{align*}
			      \h(\FPhi{\om^{\omega^\alpha}}) = \o(\FPhi{\om^{\omega^\alpha}}) = \sum_{\beta <
				      \omega^\alpha} \omega^\beta \\
			      \w(\FPhi{\om^{\omega^\alpha}}) = \sup_{\beta < \omega^\alpha} \omega^\beta
		      \end{align*}

		      The second expression clearly evaluates to
		      $\omega^{\omega^\alpha}$ as desired.
		      Since $\om^\alpha$ is a limit ordinal,
		      \Cref{lem:sum-equals-sup-ordinals} immediately gives
		      $\h(\FPhi{\om^{\om^\alpha}}) = \o(\FPhi{\om^{\om^\alpha}}) = \om^{\om^\alpha}$.

		\item For the last case: if
		      $\alpha = \om^{\alpha_1} + \cdots + \om^{\alpha_n}$ in CNF then
		      the induction hypothesis and \Cref{tab:fam:la-famille} yield:
		      \begin{equation*}
			      \h(\FPhi{\om^\alpha}) = \o(\FPhi{\om^\alpha}) =
			      \om^{\om^{\alpha_1}} \bm{\cdot} \cdots \bm{\cdot} \om^{\om^{\alpha_n}} =
			      \om^{\alpha}
		      \end{equation*}

		      For the width, $\om^{\alpha_i}$ is  \kl{additively indecomposable} for $i\in[1,n]$, hence
		      \begin{align*}
		      \w(\FPhi{\om^\alpha}) &= \om^{\om^{\alpha_1}} \odot \cdots \odot \om^{\om^{\alpha_n}}\text{ according to \Cref{tab:fam:la-famille},}\\
		      &= \om^{\om^{\alpha_1}} \bm{\cdot} \cdots \bm{\cdot} \om^{\om^{\alpha_n}} \text{ (\cref{fact:hprod-product}).}
		      \end{align*}
	\end{itemize}
\end{proof}

\end{document}